%% file: document.tex
\documentclass[]{article}
\usepackage{hyperref}
\usepackage{arxiv}  

\usepackage[T1]{fontenc}
\usepackage[utf8]{inputenc}

\usepackage{pgfplots}
\usepackage{amsmath}
\usepackage{multirow}
\usepackage{array}
\usepackage{graphicx}
\usepackage{xcolor}
\usepackage{xspace}
\usepackage{tikz}
\usepackage{caption}
\usepackage{subcaption}
\usetikzlibrary{calc,patterns,pgfplots.groupplots}

\usepackage{caption}
\usepackage{cite}
\usepackage{float}
\usepackage[ruled,vlined]{algorithm2e}

\newcommand{\MDP}{MDP\xspace}

\newcommand{\TIME}{\ensuremath{t}\xspace}
\newcommand{\NODES}{\ensuremath{\mathcal{N}}\xspace}

\newcommand{\SUBSLOTS}{\ensuremath{M}\xspace}
\newcommand{\SUBSLOT}{\ensuremath{m}\xspace}
\newcommand{\QDECISION}{\texttt{QDecision}\xspace}
\newcommand{\QEVALUATION}{\texttt{QEvaluation}\xspace}

\newcommand{\GTSREQUEST}{GTS-request\xspace}
\newcommand{\GTSRESPONSE}{GTS-response\xspace}
\newcommand{\GTSNOTIFY}{GTS-notify\xspace}
\newcommand{\RL}{RL\xspace}
\newcommand{\QL}{QL\xspace}
\newcommand{\lr}{\ensuremath{\alpha}\xspace}

\newcommand{\STATE}{\ensuremath{\mathcal{S}}\xspace}
\newcommand{\ACTION}{\ensuremath{a}\xspace}
\newcommand{\ACTIONS}{\ensuremath{\mathcal{A}}\xspace}
\newcommand{\REWARD}{\ensuremath{\mathcal{R}}\xspace}

\newcommand{\QSEND}{\textit{QSend}\xspace}
\newcommand{\QCCA}{\textit{QCCA}\xspace}
\newcommand{\QBACKOFF}{\textit{QBackoff}\xspace}

\newcommand{\CAUTIOUS}{cautious startup\xspace}
\newcommand{\PAREX}{parameter-based exploration\xspace}

\title{QMA: A Resource-efficient, Q-learning-based Multiple Access Scheme for the IIoT} 
\newcommand{\shorttitle}{QMA: A Ressource-efficient, Q-Learning-based Multiple Access Scheme for the IIoT}
\author{Florian Meyer, Volker 
	Turau\\\\Institute of Telematics\\Hamburg University of Technology}

\begin{document}
	\def\FIGWIDTH{\textwidth}
    \def\FIGHEIGHT{5cm}
	\definecolor{color0}{RGB}{117, 189, 136}
	\definecolor{color1}{RGB}{212, 114, 114}
	\definecolor{color2}{RGB}{157, 167, 209}
	\definecolor{color3}{RGB}{242, 179, 102}
	
	\maketitle
	
	\begin{abstract}
		Contention-based wireless channel access methods like CSMA and ALOHA paved the way for the rise of the Internet of Things in industrial applications (IIoT). However, to cope with increasing demands for reliability and throughput, several mostly TDMA-based protocols like IEEE 802.15.4 and its extensions were proposed. Nonetheless, many of these IIoT-protocols still require contention-based communication, e.g., for slot allocation and broadcast transmission. In many cases, subtle but hidden patterns characterize this secondary traffic. 
		Present contention-based protocols are unaware of these hidden patterns and can therefore not exploit this information.
		Especially in dense networks, they often do not provide sufficient reliability for primary traffic, e.g., they are unable to allocate transmission slots in time.
		In this paper, we propose QMA, a contention-based multiple access scheme based on Q-learning, which dynamically adapts transmission times to avoid collisions by learning patterns in the contention-based traffic. QMA is designed to be resource-efficient and targets small embedded devices. We show that QMA solves the hidden node problem without the additional overhead of RTS / CTS messages and verify the behaviour of QMA in the FIT IoT-LAB testbed. Finally,  QMA's scalability is studied by simulation, where it is used for GTS allocation in IEEE 802.15.4 DSME. Results show that QMA considerably increases reliability and throughput in comparison to CSMA/CA, especially in networks with a high load. 
	\end{abstract}
	
	\section{Introduction}
	Over the last few years, wireless communication started replacing wired communication in industrial applications due to its ease of deployment and reduced setup cost. This trend is now known as the \textit{Industrial Internet of Things} (IIoT). Many established \textit{Media Access Control} (MAC) protocols for the IoT, like S-MAC and T-MAC \cite{demirkol2006mac}, employ pure contention-based channel access and therefore struggle to fulfil the demands for reliability and timeliness imposed by industrial scenarios \cite{Telematik_Kauer_2019_Diss}. Thus, the IEEE 802.15.4 standard was extended by some sub-standards which feature TDMA-based channel access to address the challenges of the IIoT \cite{dsme_standard}. Nonetheless, contention-based communication is still required for secondary traffic, e.g., for management traffic like slot (de)allocations. For example, IEEE 802.15.4's \textit{Deterministic and Synchronous Multi-channel Extension} (DSME) employs a phase with CSMA/CA for broadcasts and for the distributed allocation of so-called \textit{Guaranteed Time Slots} (GTS) - slots spread over time and frequency offering exclusive channel access for a pair of nodes. GTS (de)allocation is based on a three-way handshake during the contention-based period. Fluctuating primary traffic therefore causes many (de)allocation messages, i.e., a high secondary traffic load during this period. Hence, using pure CSMA/CA will lead to many collisions which in turn delays slot allocation and ultimately decreases the throughput of the primary traffic. Thus, to better cope with fluctuating traffic, a more reliable channel access scheme for the contention-based period is required. 
	
	Managing contention-based channel access is difficult because no node possesses information about other nodes in the network. However, often hidden patterns introduced by primary traffic govern the secondary traffic. A good example is the 3-way handshake in DSME, which is initiated by transmitting a GTS-request and answered with a  GTS-response by the communication partner. If primary traffic is fluctuating and many (de)allocations occur, nodes can learn that it is beneficial to give the communication partner time for transmitting the GTS-response instead of transmitting another packet right away. Another example are routing protocols that transmit periodic broadcasts for route discovery such as AODV or GPSR. Of course, such patterns also emerge in primary traffic, but the scenarios are more restricted. It is impossible to identify hidden patterns when nodes are subject to frequently and spontaneously changing traffic conditions, i.e., when the transmission process is random. On the other hand, the induced secondary traffic might still inhibit hidden patterns as discussed above. Present contention-based channel access schemes like CSMA/CA do not utilize information from these patterns as it is hard to extract.
	
	Machine learning (ML) offers a promising solution to find such hidden patterns efficiently and to deduce the state of the network, e.g., traffic patterns of other nodes, from local observations \cite{lauer2000algorithm}. Our goal is a contention-based multiple access scheme that continuously adapts to changing internal and external constraints. Internal constraints are caused by temporally and spatially fluctuating primary traffic, while external constraints are due to temporally and spatially varying influences from outside of the network. This goal excludes offline ML techniques where a model is trained with a large-scale dataset and then deployed in nodes. That is because it is unfeasible to recollect and reorganize the data and retrain the model whenever the environment changes. On the other hand, models that can react to many different environmental conditions, e.g., artificial neural networks, tend to be too resource-intensive for embedded devices in wireless sensor networks. Traditional adaptability is achieved by feedback control where an algorithm tunes key parameters according to the operating conditions and prediction results. From the perspective of control, reinforcement learning (RL) is closely related to both optimal control and adaptive control. RL is a method for solving optimization problems that involve an agent interacting with its environment in a sequential style and modifying its actions, or decision policies, based on the reward received in response to its action \cite{watkins1992q}. RL refers to a class of methods that enable the design of adaptive controllers that learn online, in real-time, and improve solutions to control problems.
	
	There are plenty of works proposing MAC protocols based on RL which learn whether a transmission is likely to be successful at a given time or not \cite{liu2006rl, lee2020corl, bayat2018multi}. Most of these protocols abstract time into frames, which are further subdivided into slots. At first, every node selects a random transmission slot in every frame and gradually learns which slots are better for transmission. This way, cooperation between the nodes emerges. Nevertheless, most works only allow transmissions with a fixed expected interval and thus cannot learn hidden patterns in the primary or secondary traffic \cite{lee2020corl,chu2012aloha,chu2015application}. Additionally, some works utilize deep RL, which is not appropriate for small embedded devices \cite{aihara2019q, kim2019learning}. 
	
	We propose QMA, a Q-learning based multiple access scheme designed to increase reliability and throughput of contention-based medium access. For this, QMA learns hidden patterns in the contention-based traffic, allowing it to assess the probability of successfully sending a packet at a given time.
	Time is divided into so-called subslots, but the transmission interval is not fixed, allowing for strongly diverging traffic rates at different nodes and times. In the extreme case, a single node can utilize all subslots if required. QMA is based on a distributed Q-learning algorithm for cooperative multi-agent systems first proposed in \cite{lauer2000algorithm}. However, we extend the proposed algorithm to stochastic environments, allowing nodes to dynamically learn hidden patterns in the contention-based traffic and hence which transmission subslots are good and which are likely to result in a collision. Thereby, QMA targets embedded, resource-restricted devices and is designed with low memory and computational overhead in mind. QMA works for primary and secondary traffic, but it must exhibit a learnable, albeit possibly unknown, pattern, i.e., transmissions cannot be completely random. For random traffic, QMA would allocate slots according to the average traffic requirements of a node. 
	
	To enhance the quality of Q-learning in frequently changing environments, we propose parameter-based exploration. The problem with undirected exploration methods like $\epsilon$-greedy is that exploration is only done once so that a node cannot adapt to changes in the network over time, or that exploration is too slow \cite{mcfarlane2018survey}. Directed exploration strategies, on the other hand, impose a memory overhead. Parameter-based exploration solves this issue by defining an exploration function based on local information, e.g., the queue level. 
	
	To validate QMA, we first show its ability to solve the hidden node problem without the overhead of additional RTS / CTS messages. Additionally, we verify QMA in a realistic environment - the FIT IoT-Lab \cite{adjih_iotlab}. Here, experiments are conducted in a small tree topology and a dense star topology to investigate typical use-cases. At last, the QMA's scalability is shown by simulation in the discrete event simulator OMNeT++. Here, we apply QMA for slot-allocation in DSME. Results show that QMA achieves the same packet delivery ratio (PDR) transmitting 50 packets/s as CSMA/CA transmitting only 10 packets/s in a hidden node scenario with 3 nodes. For management traffic in IEEE 802.15.4 DSME, QMA achieves a higher PDR than CSMA/CA and manages to (de)allocate up to twice more TDMA-slots per second.
	
	\section{Related Work}
	Communication in wireless ad-hoc networks, due to its requirement of adapting to constant environmental changes, provides an optimal playground for the application of reinforcement learning. Over the past few years, a lot of work has been done optimizing selected aspects of the MAC-layer \cite{wilhelmi2017implications, ma2019reinforcement, kim2019learning} up to replacing the whole MAC-layer by learning agents \cite{liu2006rl, lee2020corl, bayat2018multi}.
	
	Liu and Elhanany propose \textit{RL-MAC} \cite{liu2006rl}, a reinforcement learning-based MAC-protocol adapting a node's duty cycle, i.e., active and sleeping times, to reduce energy consumption. The core idea is to allocate time slots for the active time, allowing a node to exchange messages with its neighbours. Afterwards, nodes transition into sleep mode. Employing Q-learning, nodes learn to maximize the ratio of the effective transmission and reception time to total active time and throughput. Liu and Elhanany compare their approach to S-MAC and T-MAC in different scenarios and, e.g., show that RL-MAC achieves up to 4 times the throughput of S-MAC in a star topology with 5 nodes while using slightly less energy. Even though their approach allows for asymmetric transmission rates at different nodes, communication during active times is still based on CSMA with RTS / CTS messages. Thus, it is random and prone to collisions during high-load traffic scenarios. It is not possible to learn hidden traffic patterns using RL-MAC.
	
	Another approach is given in \cite{lee2020corl} in the form of \textit{CoRL}, a collaborative reinforcement learning-based MAC-protocol. CoRL is based on ALOHA-Q \cite{chu2015application}. CoRL divides time into frames which are further subdivided into distinct time slots. Learning is done through stateless Q-learning, i.e., the state is invariant over time. Thus, Q-values are directly calculated for every action, which correspond to choosing a respective slot of the current frame for transmission. Additionally, a collaborative Q-value is incorporated. This design, however, severely limits the flexibility of CoRL because a node must select a single slot per frame. Compared to QMA, asymmetric traffic patterns and adaptability to hidden traffic patterns is not possible. On the other hand, CoRL requires less memory because of stateless Q-Learning. For evaluation, CoRL is compared to ALOHA-Q and it is shown that introducing a collaborative Q-value greatly reduces the time until a collision-free schedule is found. For example, in a network with 50 nodes, ALOHA-Q requires about 574 frames for convergence while CoRL only requires 420.
	
	Chu et al. propose an extension of ALOHA-Q \cite{chu2012aloha}, ALOHA-QIR, which applies Q-learning to slotted ALOHA but employs \textit{Information Receiving} (IR) to increase energy efficiency. IR appends a number $m$ to every transmitted packet which indicates the number of future frames for which the transmitter is going to use a selected slot. This allows the receiver to efficiently manage its transceiver and learn, over $m$ Q-value updates, that the slot is good for receiving packets. Additionally, they employ \textit{ping}-packets to allow nodes with low traffic to signal that a slot is still active, even when no data packet is available for transmission. Such energy-saving mechanisms are currently not considered in QMA. ALOHA-QIR allows for asymmetric traffic rates through ping-packets but compared to QMA, additional messages are required. Additionally, every node has to select at least one slot for transmission in ALOHA-QIR even when no transmission has to be conducted. Chu et al. evaluate ALOHA-QIR by simulation in a data-collection scenario with 50 randomly deployed nodes. Results show that ALOHA-QIR doubles throughput for high data rates when compared to slotted ALOHA while halving energy-consumption for any data rate.
	
	At last, Bayat-Yeganeh et al. propose a multi-state Q-learning based CSMA MAC protocol \cite{bayat2018multi}. They conceive different transmission behaviours based on a Q-agent's learned state: a simple, selfish, cautious, shapely, and holistic Q-learner. States are obtained by observing the wireless medium for varying durations and encoding the number of sensed collisions and free slots in the states. The idea of observing the channel is similar to QMA but the behaviour of a node is not learned locally but through a cooperative Q-learning algorithm. Bayat-Yeganeh et al. compare the different Q-learners with p-persistent CSMA and a random strategy to illustrate the respective behaviour of the agents. All agents behave quite differently but generally achieve higher throughput than p-persistent CSMA and the random policy in a scenario with an increasing number of nodes.
	
	Apart from the mentioned MAC-protocols, many works deal with general resource allocation and coordination using RL \cite{aihara2019q, wilhelmi2017implications, ma2019reinforcement}. These algorithms can potentially be applied MAC-protocols but they often lack domain-specific mechanisms to cope with the rapidly changing nature of wireless sensor networks. Additionally, it should be noted that most of the presented works do not utilize dedicated algorithms for multi-agent Markov decision processes as proposed in \cite{lauer2000algorithm} or \cite{abtahi2008solving}. In \cite{abtahi2008solving} a learning automaton is employed to solve the problem while Lauer and Riedmiller alter the original Q-learning algorithm to deduce local rewards from a global reward table \cite{lauer2000algorithm}. 
	
	\section{Q-Learning basics} \label{sec:q_learning}
	\textit{Q-Learning} (\QL) is a model-free, off-policy \textit{reinforcement learning} (\RL) algorithm which optimizes a policy of taken actions using rewards. Model-free means that no underlying model is required for learning but the algorithm finds an optimal policy by exploring a set of available actions. Exploration can be realized by taking random actions so that the desired policy does not have to be followed during training, i.e., off-policy learning \cite{watkins1992q}. Opposed to that, in model-based RL actions are learned - or \textit{planned} - according to known system dynamics and state transitions. 
	
\begin{figure}[h]
		\centering
		\includegraphics{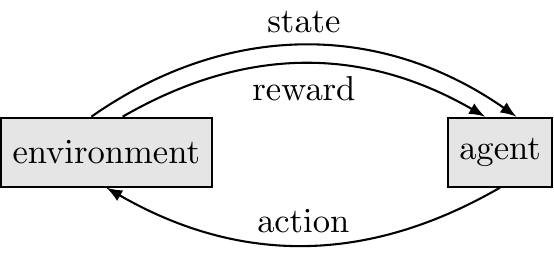}
		\caption{Interaction of an agent with its environment in \RL.}
		\label{fig:reinforcement_learning}
	\end{figure}

	The basic idea of \RL is that an agent interacts with its environment. For this, it observes state $\STATE_t$ of the environment at time $t$ and executes an action $\ACTION_t \in \ACTIONS_t$ in response to it, changing the state of the environment. $\ACTIONS_t$ is the set of available actions at time $t$. The agent can then observe the next state $\STATE_{t+1}$ and receives a reward $\REWARD_t$ for the chosen action. The goal is to maximize the total reward, i.e., the sum of rewards attained in $\STATE_t$ and all future states $\STATE_{t+i}$. This way, the agent can learn an optimal policy of actions for any finite Markov decision process (\MDP) \cite{melo2001convergence}. The interaction between the agent and the environment is also depicted in Fig.~\ref{fig:reinforcement_learning}. 
	
	\QL uses so-called Q-values, associated with $\STATE_t$ and $\ACTION_t$, to express the quality of executing $\ACTION_t$ in state $\STATE_t$. To incorporate future rewards while updating Q-values, \QL relies on a weighted form of the Bellman equation which can be expressed as 
	\begin{align}
		Q(\STATE_t, \ACTION_t) &\leftarrow (1-\lr) Q(\STATE_t, \ACTION_t) + \lr (\REWARD_t + \gamma \max_a Q(\STATE_{t+1}, a)) \label{eq:q_update}. 
	\end{align}
	
	Here, $\lr$ is the learning rate, determining the influence of a newly calculated Q-value, and $\gamma$ is a discount factor, 
	determining the influence of future rewards \cite{watkins1992q}. In other words, 
	$\gamma$ expresses how far the agent looks into the future. 
	For $\gamma=0$ it only considers the current reward, while it 
	considers all future rewards for $\gamma = 1$. For systems without a final state, $\gamma < 1$ is chosen to ensure that $Q(\STATE_t, \ACTION_t)$ does not become infinite by summation of all future rewards. 
	
	\subsection{Cooperative multi-agent Q-learning} \label{sec:coop_q_laerning}
	In this work, agents are represented by network nodes, which are supposed to learn globally optimized behaviour independently. Thus, an adapted Q-learning method for multi-agent systems is required. Lauer and Riedmiller proposed an enhanced method for Q-Learning in cooperative multi-agent systems \cite{lauer2000algorithm}. The idea is that complex interactions between multiple agents cannot be modelled by local MDPs at each node. Thus, they utilize a Q-Table, which models rewards for all combinations of actions in the network. An example is given by Tbl.~\ref{tbl:dist_q_learning_ex} for two agents with two actions.
	
	\begin{table}[h]
		\centering
		\begin{tabular}{cc||c|c}
			\multicolumn{1}{c}{Global} & & \multicolumn{2}{c}{agent 0} \\
			\multicolumn{1}{c}{Q-Table} & & a' & a'' \\
			\hline \hline
			\multirow{2}{1.2cm}{agent 1}  & a' & 1 & -1 \\
			& a'' & -1 & 10 \\
		\end{tabular}
		\hspace{1cm}
		\begin{tabular}{c|c}
			\multicolumn{2}{c}{agent 0} \\
			\multicolumn{2}{c}{(local)} \\
			a' & a'' \\
			\hline \hline
			1 & 10 \\
		\end{tabular}
		\hspace{1cm}
		\begin{tabular}{c|c}
			\multicolumn{2}{c}{agent 1} \\
			\multicolumn{2}{c}{(local)} \\
			a' & a'' \\
			\hline \hline
			1 & 10 \\
		\end{tabular}
		
		\caption{Example of a global Q-table for two agents with two actions and the respective local Q-tables.}
		\label{tbl:dist_q_learning_ex}
	\end{table}
	
	Under the assumption that all agents behave cooperatively - i.e. they try to choose actions which maximize the reward from the conceptual global Q-table, every agent updates its local Q-table per state $\STATE_t$ and action $\ACTION_t \in \ACTIONS_t$ according to 
	\begin{align}
		Q(\STATE_t, \ACTION_t) \leftarrow \max\{Q(\STATE_t, \ACTION_t), \REWARD_t + \gamma \max_a Q(\STATE_{t+1}, a)\}. \label{eq:dist_q_update} 
	\end{align}
	Here, $Q(\STATE_t, \ACTION_t)$ is only updated if the newly calculated Q-value is larger than $Q(\STATE_t, \ACTION_t)$. This ensures an optimal policy after all possible actions have been explored. For example, agent 1 could choose action a''. Depending on agent 0, agent 1 experiences two different rewards -1 and 10. It only stores a reward of 10 for action a'', as given by Eq.~\ref{eq:dist_q_update}, under the assumption that agent 0 also chooses action a'' to maximize the global reward. Thus, the agents achieve cooperation. 
	
	According to Lauer and Riedmiller, a problem arises if multiple policies lead to an optimal reward as shown in Tbl.~\ref{tbl:dist_q_learning_ex2}. The combination of actions (a', a') and (a'', a'') both yield a reward of 10 and in compliance with Eq.~\ref{eq:dist_q_update}, both agents update their local Q-values to $Q(\STATE_t, a') = 10$ and $Q(\STATE_t, a'') = 10$. Therefore, agent 0 could choose action a' while agent 1 selects action a'' as both actions yield the same reward. The combination (a', a''), however, only yields a reward of -1. Thus, the agents fail to agree on an optimal policy. 
	
	\begin{table}[h]
		\centering
		\begin{tabular}{cc||c|c}
			\multicolumn{1}{c}{Global} & & \multicolumn{2}{c}{agent 0} \\
			\multicolumn{1}{c}{Q-Table} & & a' & a'' \\
			\hline \hline
			\multirow{2}{1.2cm}{agent 1}  & a' & 10 & -1 \\
			& a'' & -1 & 10 \\
		\end{tabular}
		\hspace{1cm}
		\begin{tabular}{c|c}
			\multicolumn{2}{c}{agent 0} \\
			\multicolumn{2}{c}{(local)} \\
			a' & a'' \\
			\hline \hline
			10 & 10 \\
		\end{tabular}
		\hspace{1cm}
		\begin{tabular}{c|c}
			\multicolumn{2}{c}{agent 1} \\
			\multicolumn{2}{c}{(local)} \\
			a' & a'' \\
			\hline \hline
			10 & 10 \\
		\end{tabular}
		
		\caption{Example of a global Q-table for two agents with two actions and the respective local Q-tables.}
		\label{tbl:dist_q_learning_ex2}
	\end{table}
	
	A solution is to maintain the optimal policy $\pi(\STATE_t)$ per state as a separate, additional table. Policy $\pi(\STATE_t)$ is learned in parallel to local Q-values as 
	\begin{align}
		\pi(\STATE_t) \leftarrow
		\begin{cases}
			\underset{a}{\text{argmax}}\ Q(\STATE_t, a) \qquad & \text{if } \underset{a}{\max} \ Q(\STATE_t, a) > \REWARD_t + \gamma \underset{a}{\max} Q(\STATE_{t+1}, a) \\ 
			\underset{a}{\text{argmax}}\ \REWARD_t + \gamma Q(\STATE_{t+1}, a) & \text{otherwise}
		\end{cases} \label{eq:policy_update}.
	\end{align}
	That means, an agent only selects a new action for $\STATE_t$ if the associated Q-value is \textit{strictly greater} than the Q-value of current policy $\pi(\STATE_t)$. This way, a switch of actions is avoided for duplicate global rewards and all agents agree on policy $\pi(\STATE_t)$ which yielded the duplicate reward first.

	\subsubsection{Stochastic environments} 
	Another problem arises in stochastic environments, i.e., when actions do not have a deterministic outcome. Consider two agents competing for access to a shared resource, where action a' acquires the resource for the current state and action a'' waits until the next state. The resource cannot be used by both agents at the same time. Table~\ref{tbl:dist_q_learning_stochastic} summarizes the global rewards. 
	
	\begin{table}[h]
		\centering
		\begin{tabular}{cc||c|c}
			\multicolumn{1}{c}{Global} & & \multicolumn{2}{c}{agent 0} \\
			\multicolumn{1}{c}{Q-Table} & & a' & a'' \\
			\hline \hline
			\multirow{2}{1.2cm}{agent 1}  & a' & -1 & 1 \\
			& a'' & 1 & 0 \\
		\end{tabular}
		\hspace{1cm}
		\begin{tabular}{c|c}
			\multicolumn{2}{c}{agent 0} \\
			\multicolumn{2}{c}{(local)} \\
			a' & a'' \\
			\hline \hline
			1 & 1 \\
		\end{tabular}
		\hspace{1cm}
		\begin{tabular}{c|c}
			\multicolumn{2}{c}{agent 1} \\
			\multicolumn{2}{c}{(local)} \\
			a' & a'' \\
			\hline \hline
			1 & 1 \\
		\end{tabular}
		
		\caption{Global rewards for resource acquisition of two agents and according local Q-tables.}
		\label{tbl:dist_q_learning_stochastic}
	\end{table}
	
	As one can see, a punishment is given if both agents try to acquire the shared resource in the same state and a reward if only one of the agents acquires the resource. However, there is a small probability that an agent chooses a' but does not acquire the resource as it is not needed by the agent in the current state. If the other agent also executes a' and acquires the resource, it experiences the full reward, thinking that the first agent chose a''. This might happen vice-versa and will eventually result in a collision during the acquisition. As both agents only store the maximum experienced reward per action, according to Eq.~\ref{eq:dist_q_update}, they are stuck with action a' which will result in many failed acquisitions over time. Lauer and Riedmiller mention this problem but do not propose a solution \cite{lauer2000algorithm}. 
	
	We offer a solution to the problem by introducing a small penalty $\xi$ for Q-value updates, s.t. 
	\begin{align}
		Q(\STATE_t, \ACTION_t) \leftarrow \max\{Q(\STATE_t, \ACTION_t) - \xi, \REWARD_t + \gamma \max_a Q(\STATE_{t+1}, a)\}. 
	\end{align}
	This way, $Q(\STATE_t, \ACTION_t$) is gradually reduced if a maximum reward was only experience once. On the other hand, $Q(\STATE_t, \ACTION_t$) is reset to the maximum Q-value every other update if the maximum reward is continuously experienced.  In other words: $\xi$ only affects fluctuating Q-values because stable and optimal Q-values are reupdated to their original value once they have been decremented. Notice that a penalty factor would not be appropriate, as it only pulls the Q-values towards zero but would not change the policy if an agent received a positive reward once but then receives punishments repeatedly. At last, learning rate \lr is not considered in \cite{lauer2000algorithm}. Similar to Eq.~\ref{eq:q_update}, it can be incorporated as 
	\begin{align}
		Q(\STATE_t, \ACTION_t) \leftarrow \max\{Q(\STATE_t, \ACTION_t) - \xi, (1-\lr)Q(\STATE_t, \ACTION_t) + \lr(\REWARD_t + \gamma \max_a Q(\STATE_{t+1}, a))\}. \label{eq:q_sub_update} 
	\end{align}
	
	\subsection{Q-value Representation} 
	One of the main challenges of \QL is representing the Q-values 
	for every state-action pair. In theory, any linear function approximator can be exploited for this without influencing the stability of the algorithm. It is often done for problems with a large or continuous action space to reduce the number of states and thus memory usage and the number of required training steps. The idea is to approximate Q-values 
	$Q(\STATE_t, \ACTION_t)$ by carefully selecting features and 
	updating the approximation using gradients. However, 
	execution can still be quite resource-intensive, e.g., when a 
	linear combination of many features is used.  
	
	Therefore, we employ a simple table with $|\STATE_t \times 
	\ACTIONS_t|$ values. Training is as simple as writing a value 
	$Q(\STATE_t, \ACTION_t)$ to the designated cell. Thus, at maximum two 
	multiplications, three additions and $|\ACTIONS_t|+1$ array lookups are 
	required for each training step, using Eq.~\ref{eq:q_sub_update}. It is possible, e.g., by selecting 
	$\lr = 0.5$ and integer rewards, to replace multiplication by the learning rate with an efficient right shift by one. This allows for the execution on resource-restricted and embedded devices without a floating-point unit.
	
	At last, it should be said that there has been a lot of effort to enable Q-Learning with special non-linear function approximators - \textit{deep neural networks } (DNNs). 
	Training DNNs through backpropagation is quite resource-intensive, rendering them unsuitable for online-training on embedded devices. Additionally, storing weights consumes plenty of memory and inference involves many vector multiplications. Therefore, only small, pre-trained 
	DNNs can be utilized and DNNs are not considered for this work. 
	
	\section{QMA: Q-learning-based multiple access} 
	\label{sec:self_synchronization}
	The core idea of QMA is to utilize Q-learning to determine at which times transmission is beneficial and at which times it is likely to result in a collision. For this, we employ a mechanism similar to slotted CSMA/CA, where time is divided into \SUBSLOTS subslots $\SUBSLOT_0,\dots,\SUBSLOT_{\SUBSLOTS-1}$, discretizing time to keep the state space small. For application in DSME, 8 CAP slots are further subdivided into 54 subslots. During this time, the transceiver is turned on to guarantee compatibility with CSMA/CA.  
	
	QMA's states only consist of the current subslot id $\SUBSLOT_t$, allowing nodes to synchronize their transmission slots solely based on time information. At first, it might seem beneficial to include the queue level, number of overheard messages, or number of successive idle slots in the state. However, several experiments have shown that these additional inputs withhold the node from building a collision-free schedule. For example, nodes learn to transmit packets when their queue is full but do not care if the according subslot is already used by another node. Instead, queue levels are incorporated into the exploration function, as discussed in Sect.~\ref{sec:parameter_based_exploration}.
	
	The action space of QMA is given by the set $\ACTIONS_t = \{\QBACKOFF, \QCCA, \QSEND\}$. The actions are defined as follows: \QBACKOFF lets the node backoff to the next subslot, \QCCA performs a clear channel assessment (CCA), transmits a packet on success or backs of to the next subslot on failure, and \QSEND transmits a packet immediately without assessing the channel. Separating \QCCA and \QSEND allows for priority-based transmissions, i.e., nodes can select \QSEND if they have to transmit a packet urgently. Other nodes can check, by selecting \QCCA, if there is already a high-priority transmission going on and backoff to a later subslot. 
	
\begin{figure}[ht]
		\centering
		\includegraphics{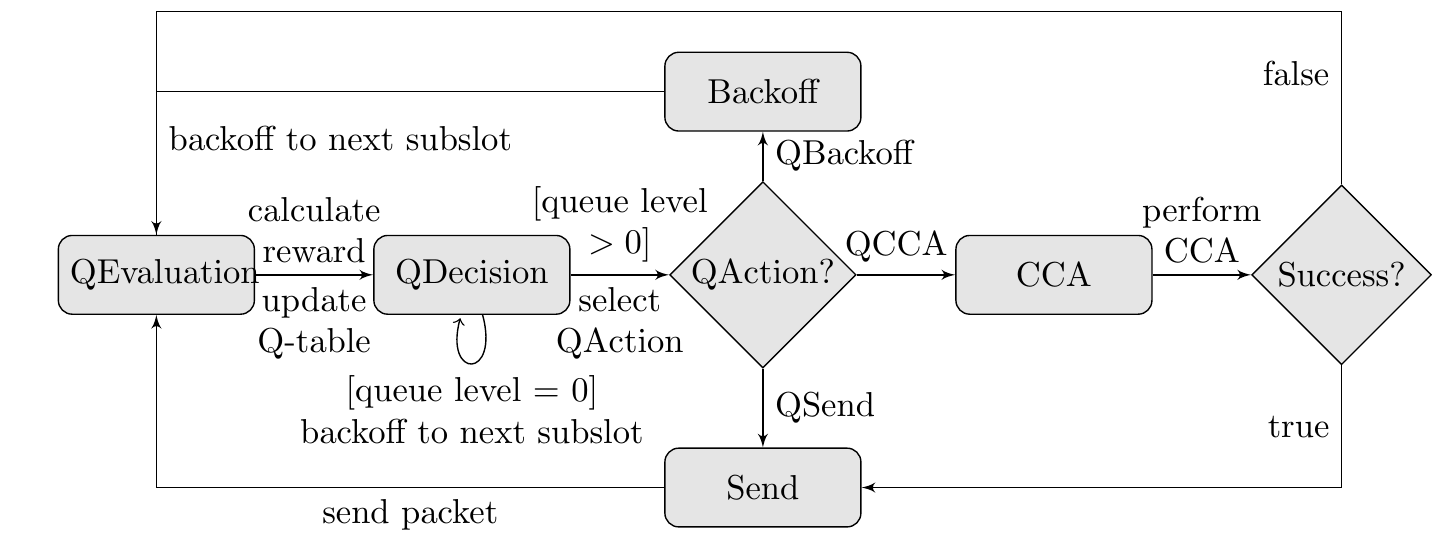}
		\caption{Flow chart for the modified CAP-layer state machine.}
		\label{fig:q_cap_flow}
	\end{figure}

	Fig.~\ref{fig:q_cap_flow} shows QMA's internal structure based on 5 states. In state \QDECISION, a node checks if a packet needs to be transmitted and selects an action $\ACTION_t \in \ACTIONS_T$ depending on $\SUBSLOT_t$. Then $\ACTION_t$ is executed and QMA transitions to state \QEVALUATION. If \QBACKOFF is executed, \QEVALUATION is guaranteed to be executed at the start of the next subslot. However, a transmission triggered by \QCCA or \QSEND can span multiple subslots, depending on the packet length. At last, $\ACTION_t$ is evaluated according to Eq.~\ref{eq:reward_backoff} to \ref{eq:reward_send}, the Q-table is updated and the next packet can be transmitted. Consequently, rewards are not directly observable after an action is performed, but, e.g., after an ACK is received. Therefore, we save the state $\STATE_t$ and action $\ACTION_t$ until the outcome of $\ACTION_t$ is known. Afterwards $Q(\STATE_t, \ACTION_t)$ is updated \cite{dulac2019challenges}. 
	
	The full algorithm is shown in Algorithm~\ref{alg:qma}. If the queue is not empty at the start of a subslot, QMA selects an action $\ACTION_t$ randomly with probability $\rho$ or according to policy $\pi$ with probability $1-\rho$. Here, $\rho$ is calculated by parameter-based exploration, which is discussed in Sect.~\ref{sec:parameter_based_exploration}.  Afterwards, action $\ACTION_t$ is executed. As explained above, $\ACTION_t$ can span multiple subslots due to processing time and, in case of action \QSEND and \QCCA, waiting time for an ACK. During the waiting time, a node is inactive and does not select further actions. At last, the Q-table is updated and policy $\pi$ is adjusted accordingly if a better action is found. Notice, that policy $\pi(\SUBSLOT_t)$ is initialized with action \QBACKOFF for every subslot $\SUBSLOT_t$ and is only changed if a strictly greater Q-value is found, as described in Sect.~\ref{sec:coop_q_laerning}. 
	
	\SetKwFor{For}{on}{do}{end}
	\begin{algorithm}
		\SetAlgoLined
		\KwIn{learning rate $\lr$, discount factor $\gamma$, penalty $\xi$, exploration rate $\rho_t$}
		initialize $\forall_{\SUBSLOT_t, \ACTION_t \in \ACTIONS_t}:\quad Q(\SUBSLOT_t, \ACTION_t) \leftarrow -\infty$  \\
		initialize $\forall_{\SUBSLOT_t}:\quad \pi(\SUBSLOT_t) \leftarrow \QBACKOFF$ \\
		\For{subslot $\SUBSLOT_t \in \{\SUBSLOT_0,\dots,m_{\SUBSLOTS-1}\}$} {
			\If{queue level > 0}{
				\eIf{random() > $\rho_t$} {	
					$\ACTION_t \leftarrow \pi(\SUBSLOT_t)$\\
				} {
					select $\ACTION_t \leftarrow$ random action \tcp*{exploration phase}
				}
				
				execute $\ACTION_t$, \textbf{wait} until $\ACTION_t$ is finished, $i \leftarrow$ passed subslots \\
				$\REWARD_t \leftarrow$ received reward \\         
				
                \If(\tcp*[f]{Eq.(\ref{eq:policy_update})}){$\ Q(\SUBSLOT_t, \ACTION_t) < (1-\lr)Q(\SUBSLOT_t, \ACTION_t) + \lr(\REWARD_t + \gamma\ \underset{a}{\text{max}}\  Q(\SUBSLOT_{t+i},a))$} {
                    $\pi(\SUBSLOT_t) \leftarrow \underset{a}{\text{argmax}}\  Q(\SUBSLOT_t, a)  $
                }
                
				$Q(\SUBSLOT_t, \ACTION_t) \leftarrow \max\{Q(\SUBSLOT_t, \ACTION_t) - \xi, (1-\lr)Q(\SUBSLOT_t, \ACTION_t) + \lr(\REWARD_t + \gamma\ \underset{a}{\text{max}}\  Q(\SUBSLOT_{t+i},a))\}$ \tcp*{ Eq.(\ref{eq:q_sub_update})}
			}
		}
		\caption{Learning multiple access with QMA}
		\label{alg:qma}
	\end{algorithm}    
	
	At last, it should be said that a packet is dropped after $NR$ retransmission as in CSMA/CA \cite{dsme_standard}. However, a packet is not discarded after $NB$ backoffs since QMA's main idea is to synchronize transmission times which might require several backoffs until an appropriate subslot for transmission is reached. 
	
	\subsection{Reward Function Design}	
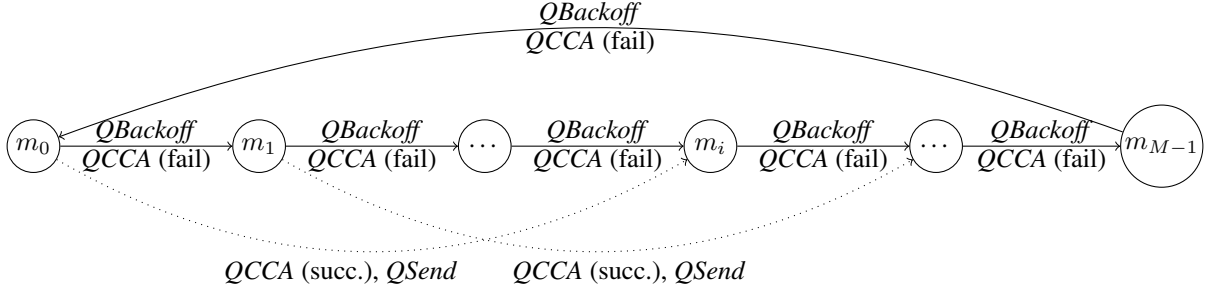
\begin{figure}[h]
		\centering
		\begin{tikzpicture}[node distance=3cm, inner sep=1pt, minimum width=.7cm]
			\node[draw=black, circle] at (0,0) (s00) {$\SUBSLOT_{0}$};
			\node[draw=black, circle, right of=s00] (s01)  {$\SUBSLOT_{1}$};
			\node[draw=black, circle, right of=s01] (s02)  {\dots};
			\node[draw=black, circle, right of=s02] (s03)  {$\SUBSLOT_{i}$};
			\node[draw=black, circle, right of=s03] (s04)  {\dots};
			\node[draw=black, circle, right of=s04] (s05)  {$\SUBSLOT_{\SUBSLOTS-1}$};
			
			\path[draw=black,->] (s00) edge node[align=center] {\QBACKOFF\\\QCCA (fail)} (s01);
			\path[draw=black,->] (s01) edge node[align=center] {\QBACKOFF\\\QCCA (fail)} (s02);
			\path[draw=black,->] (s02) edge node[align=center] {\QBACKOFF\\\QCCA (fail)} (s03);
			\path[draw=black,->] (s03) edge node[align=center] {\QBACKOFF\\\QCCA (fail)} (s04);
			\path[draw=black,->] (s04) edge node[align=center] {\QBACKOFF\\\QCCA (fail)} (s05);
			\path[draw=black,->] (s05) edge[bend right=20] node[align=center] {\QBACKOFF\\\QCCA (fail)} (s00);
			
			\path[draw=black,->,dotted] (s00) edge[below,bend right,pos=0.45] node[yshift=-0.1cm] {\QCCA (succ.), \QSEND} (s03);
			\path[draw=black,->,dotted] (s01) edge[below,bend right,pos=0.55] node[yshift=-0.1cm] {\QCCA (succ.), \QSEND} (s04);
		\end{tikzpicture}
		\caption{Exemplary multi-agent hidden Markov 
			decision process. Hidden interactions are the result of concurrent actions of nodes with an overlapping transmission range, they are symbolized by dotted lines.}
		\label{fig:multi_agent_markov_decision_process}
	\end{figure}

	The goal of the multi-agent Q-learning algorithm is to find a globally optimized policy using local information. Thus, it is necessary to define rewards for the interaction of different nodes, i.e., rewards for the set of actions $\{a_0,\dots,a_{\NODES-1}\}$ of nodes $n_0,\dots,n_{\NODES-1}$ at time \TIME. The interactions are illustrated by the hidden Markov decision process in Fig.~\ref{fig:multi_agent_markov_decision_process}. As one can see, actions \QBACKOFF and \QCCA, after a failed CCA, backoff to the next subslot. Thus, there is no interaction with other nodes in the network. After subslot $\SUBSLOT_{\SUBSLOTS-1}$ the agent goes back to the first subslot $\SUBSLOT_0$. On the other hand, actions \QSEND and \QCCA, after a successful CCA, interact with the network by transmitting a packet and possibly interfering with other transmissions. As shown, a transmission can span multiple subslots, depending on the packet's length. Obviously, in an optimal policy, a single node performs action \QSEND while the other nodes choose action \QBACKOFF so that there is no collision.	
	
	Without loss of generality and for sake of simplicity, we 
	consider the interaction of 3 Q-Agents with actions 
	\QBACKOFF, \QCCA and \QSEND. This is sufficient because there is no difference in a collision of 2 or $n$ packets and all observing nodes behave the same. Local rewards are 
	necessary because a single node is not able to observe 
	the global state of the network, i.e., the actions executed at other nodes. 
	Table~\ref{tbl:local_rewards} shows the local rewards 
	for all possible interactions between the nodes and the 
	resulting global reward. 
	
	\begin{table}[h]
		\centering
		\begin{tabular}{c|>{\centering\arraybackslash}p{2cm}|>{\centering\arraybackslash}p{2.5cm}|c} 
			&Actions \newline $a_0$ / $a_1$ / $a_2$ & Local rewards \newline $r_0$ / $r_1$ / $r_2$ & Global reward \REWARD \\ \hline 
			
			\multirow{3}{2.7cm}{Successful transmission}  
			& B S B & 2 / 4 / 2 & 8 \\
			&B C B & 2 / 3 / 2 & 7 \\
			&C S C & 1 / 4 / 1 & 6 \\ \hline
			\multirow{1}{2.7cm}{No transmission} 
			&B B B & 0 / 0 / 0 & 0 \\  \hline
			\multirow{5}{2.7cm}{Failed transmission}
			&C B C & -2 / 0 / -2 & -4 \\
			&S B C & -3 / 0 / -3 & -6 \\
			&C C C & -2 / -2 / -2 & -6 \\
			&S C S & -3 / 1 / -3 & -5 \\
			&S S S & -3 / -3 / -3 & -9 \\
		\end{tabular}
		\caption{Local rewards and conceptual global rewards for all interactions of Q-Agents with actions \QBACKOFF (B), \QCCA (C) and \QSEND (S).}
		\label{tbl:local_rewards}
	\end{table}
	
	Here, it can be seen that a reward is given individually for every action in response to the observed state of the channel. In particular, the local reward $\REWARD_L$ for the actions \QBACKOFF, \QCCA and \QSEND is given by: 
	\begin{itemize}
		\item \QBACKOFF:
		\begin{align}
			\REWARD_L =
			\begin{cases}
				2  \qquad & \text{if DATA or ACK packet overheard}\\
				0  \qquad & \text{otherwise}
			\end{cases}
			\label{eq:reward_backoff}
		\end{align}
		\item \QCCA:
		\begin{align}
			\REWARD_L = 
			\begin{cases}
				3  	  \qquad & \text{if CCA success and TX success}\\
				-2     \qquad &\text{if CCA success and TX failed}\\
				1     \qquad &\text{if CCA failed}
			\end{cases}
			\label{eq:reward_cca}
		\end{align}
		\item \QSEND:
		\begin{align}
			\REWARD_L = 
			\begin{cases}
				4    \qquad & \text{if TX success} \qquad\qquad\qquad\mkern18mu \\
				-3  \qquad & \text{if TX failed}\\
			\end{cases}
			\label{eq:reward_send}
		\end{align}
	\end{itemize}
	\QSEND is a high-risk, high-reward action which results in the largest reward and largest punishment if a transmission is successful or unsuccessful, respectively. Thus, the reward for a successful transmission using \QCCA is slightly lower because it implies additional time overhead and energy consumption for a CCA. The same applies to \QCCA and \QBACKOFF. If no data packet is sent, a higher reward is given for \QBACKOFF because it does not result in a potential collision. Different rewards have been experimentally evaluated, including larger differences between the rewards. However, larger differences often favour one action over another. For example, increasing the reward for a successful transmission using \QSEND to 8 results in a policy where every node executes \QSEND in every subslot. This is not desired. Therefore, the presented reward function is a careful balance between all actions supporting the desired objectives of increased throughput and reliability. It is the result of many experiments. The design of the reward function is an art in itself, there are no common blueprints known. 
	
	Initially, Q-values are initialized to $-\infty$ because Eq.~\ref{eq:q_sub_update} only updates the policy if a calculated Q-value is higher than the one in the Q-table. In practice, it is sufficient to choose a number smaller than the largest punishment. Thus, we initialize the Q-table to -10. It should be mentioned that, by observing the state of the channel and transmissions, QMA cannot distinguish between internal and external interference and computational overload of a communication partner. QMA also learns these patterns and adapts to them dynamically. 
	
	\subsection{Parameter-based exploration} \label{sec:parameter_based_exploration}
	Some works have already described the importance of exploration for the quality of Q-learning-based MAC-protocols~\cite{kosunalp2016use, lee2020corl}. One of the more popular exploration strategies is $\epsilon$-greedy. It selects random actions with probability $\epsilon>0$, where $\epsilon$ is exponentially decreased over time. Another option is to utilize a constant exploration rate. Both of these methods are not particularly suitable for dynamically changing systems. The goal of $\epsilon$-greedy is to reach an optimal stable state as fast as possible, however, $\epsilon$ is never increased again so that changes in the environment cannot be explored. Additionally, $\epsilon$-greedy initially results in many collisions because many random actions are selected \cite{alsheikh2014machine}. On the other hand, a constant exploration rate tends to be too slow, and increasing it would result in too many random actions in a stable state. 
	
\begin{figure}
		\centering
		\input{./plots/parameter_based_exploration.tex}
		\caption{Dependency of $\rho$ for parameter-based exploration on the queue level in QMA.}
		\label{fig:parameter_based_exploration}
	\end{figure}
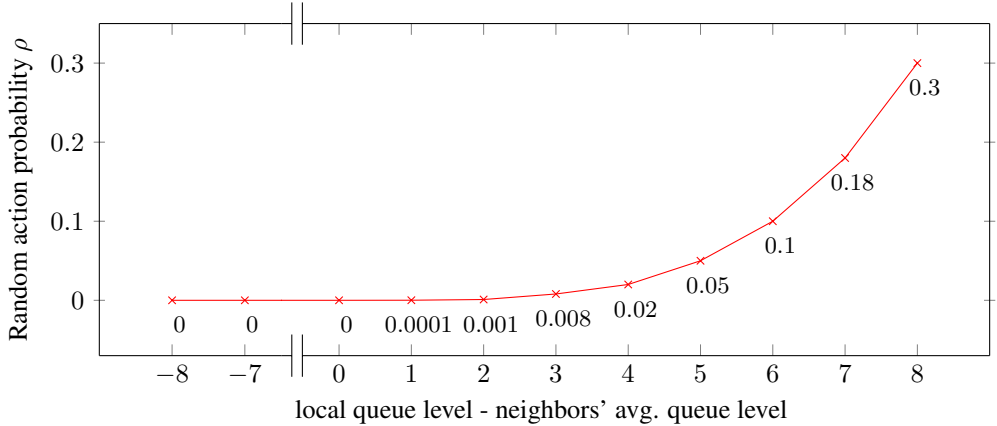

	We propose a parameter-based exploration mechanism in which actions are randomly selected with probability $\rho$, where $\rho$ increases exponentially based on given parameters, such as queue size. A stable state can usually be recognized by utilizing local information: queues are empty, as all packets are transmitted successfully. Changing a network's state, e.g., when a new node joins the network, results in many collisions because the new node first executes several random actions. In this case, queue levels increase and more random actions are executed to find a stable state, while actions are selected greedily when a stable state is reached. However, when the network is oversaturated, it is not sufficient to only look at the local queue level because all queues in the network are full. Therefore, the average queue level of all neighbouring nodes is subtracted from the local queue level to assess if the network is oversaturated or if all other nodes transmit their packets successfully. The current queue level of a neighboring nodes is piggybacked into regular data messages.  Fig.~\ref{fig:parameter_based_exploration} illustrates how $\rho$ is chosen based on the difference between the local queue level and the average queue level of all neighbours in QMA. It is not desirable to execute actions with full randomness as it would destroy an already established action schedule. Therefore, $\rho=0.3$ is used for the maximum queue level of 8 packets. Since the maximum queue length is usually quite small, $\rho$ is stored in a table and can be used efficiently by resource-restricted devices without any computational overhead. If the average queue level of all neighbouring nodes is larger than the local queue level, $\rho=0$ to prevent further exploration and give neighboring nodes a chance to allocate additional slots. 
	
	The application of parameter-based exploration is also conceivable in other contexts. For example, nodes could start exploration based on available energy and stop exploration when energy is becoming low.
	
	\subsection{Cautious startup} \label{sec:cautious_startup}
	As discussed in Sect.~\ref{sec:parameter_based_exploration}, nodes joining an already built-up network are likely to destroy a created schedule. We combat this effect with a cautious startup mechanism. For the first $\Delta$ subslot iterations after first selecting an action, a node only executes action \QBACKOFF and observes the wireless medium. Overheard packets are registered and a reward is given according to Eq.~\ref{eq:reward_backoff}. Additionally, action \QCCA and \QSEND are punished with a negative reward of -2 and -3, respectively. This way, a subslot already used for transmission by another node is indicated in the Q-table. This cannot prevent the node from randomly transmitting a packet in any such slot, but there is an initial bias that prevents the node from fully committing to the slot. That means if the node transmits a packet and there was no collision because the other node did, by chance, not have a packet to send, it receives the positive reward. However, with $\lr=0.5$ it takes multiple successful transmissions to change the policy for the slot because \QBACKOFF also received a positive reward initially.

	\section{QMA: An example} \label{sec:qma_example}
	\newcommand{\nA}{\ensuremath{n_1}\xspace}
	\newcommand{\nB}{\ensuremath{n_2}\xspace}
	\newcommand{\nC}{\ensuremath{n_3}\xspace}
	
\begin{figure}[h]
		\centering
		\includegraphics[width=\textwidth]{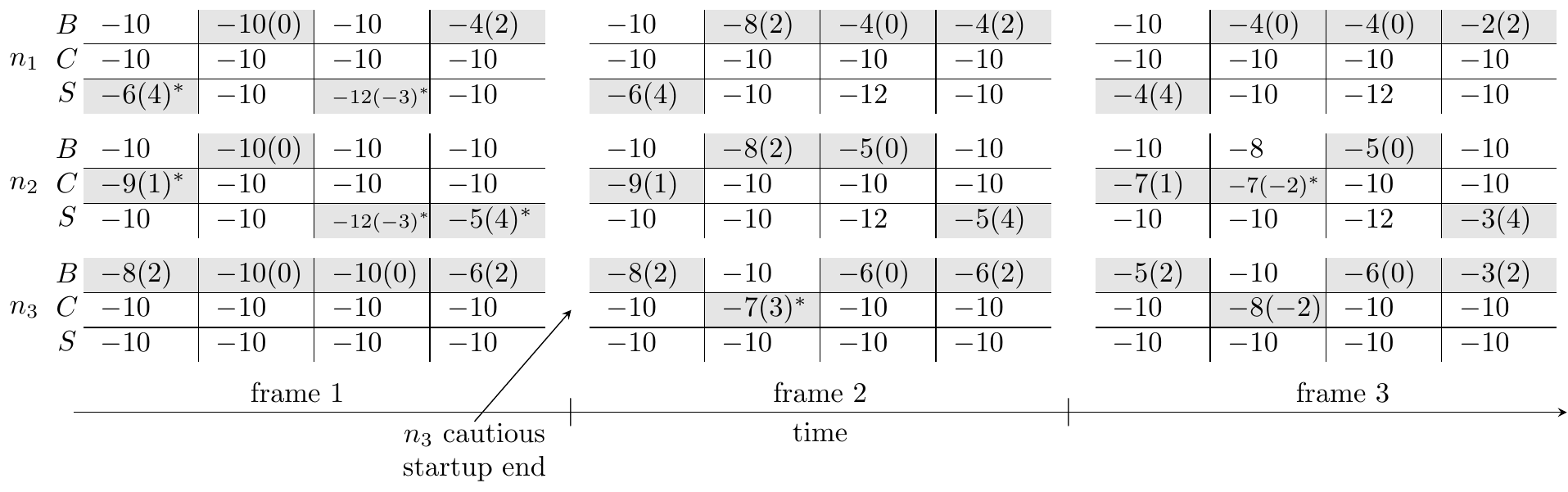}
		\caption{Example Q-tables during QMA's execution with 3 nodes and 4 subslots per CAP with $\QBACKOFF (B)$, $\QCCA (C)$ and $\QSEND (S)$. Actions taken by an agent are highlighted. The received reward for any action is given in brackets. Random actions are indicated by a star (*) and $\xi=2$.}
		\label{fig:qma_example}
	\end{figure}

	To illustrate how all of QMA's components work together, a small example with three nodes (\nA,\nB,\nC) is given in Fig.~\ref{fig:qma_example}. It shows the evolution of Q-tables with actions $\QBACKOFF (B)$, $\QCCA (C)$ and $\QSEND (S)$ over the course of three frames. For sake of simplicity, consider four subslots per CAP, where the executed action is highlighted and the respective reward is given in brackets. Random actions are indicated by a star and are triggered by \PAREX. Notice that the Q-table's state is depicted \textit{after} every frame and that Q-values are initialized to -10. Every action is executed within a single subslot and the policy $\pi$ is initialized to \QBACKOFF, as described in Algorithm~\ref{alg:qma}. Additionally, all updates are calculated with Eq.~\ref{eq:q_sub_update} and Eq.~\ref{eq:policy_update} with $\lr=1$ and $\gamma=1$, and nodes always have at least one packet in the queue for transmission. 
	
	As depicted in Fig.~\ref{fig:qma_example}, \nC is still in \CAUTIOUS during the first frame. Thus, as described in Sect.~\ref{sec:cautious_startup}, it only selects action \QBACKOFF and observes the channel. \nA and \nB, on the other hand, already started exploration, resulting in \nA executing \QSEND and \nB executing \QCCA randomly in the first subslot. \nA successfully transmits a packet and receives a reward of 4, according to Eq.~\ref{eq:reward_send}, while \nB receives a reward of 1 for a failed CCA, as stated in Eq.~\ref{eq:reward_cca}. The Q-values are updated according to Eq.~\ref{eq:q_sub_update}, where the maximum Q-value of the next state is -10 at the beginning of the first frame, as it has not been visited yet.  In subslot 4 of the first frame, the same applies, but a Q-value of the next state (subslot 1) has already been updated. Thus, the Q-value of the current subslot $Q(\STATE_t, \ACTION_t)$ is updated to -4 because $\REWARD_t = 2$ and $\max_{a} Q(S_{t+1},a) = -6$ in Eq.~\ref{eq:q_sub_update}. \nC overhears the successful transmissions in subslot 1 and 4 and receives a reward of 2 for this action, according to Eq.~\ref{eq:reward_backoff}.
	
	An interesting case occurs in subslot 3 of frame 1. Here, \nA and \nB choose \QSEND and receive a reward of -3 for a collision with respect to Eq.~\ref{eq:reward_send}. However, the Q-values are not updated to -13 but -12 because the newly calculated Q-value is not bigger than the Q-value in the Q-table. Instead, $\xi=2$ is subtracted from the Q-value in the Q-table. \nC does not overhear a packet due to the collision. One should note that the current policy is only changed if a larger Q-value is found. Thus, \nA and \nB execute \QBACKOFF in the next frame. 
	
	During the next two frames, the Q-tables start to settle and less randomness is involved in the action selection. \nC randomly selects \QCCA in subslot 2 of frame 2 and transmits a packet successfully. Therefore, every node now has one subslot for transmission. At his point, it has to be said that more exploration and consequently more collisions occur in the actual execution of QMA. However, mechanisms like \CAUTIOUS and \PAREX help to reach a stable state faster.
	
	If $\gamma < 1$, the Q-table converges to a state where most chosen actions are associated with a positive Q-value. Subslots with successful transmission are associated with large positive Q-values while failed transmissions are indicated with large negative Q-values for action \QSEND. A more detailed analysis of QMA's convergence is given in Sect.~\ref{sec:evaluation_hidden_node_convergence} and \ref{sec:evaluation_hidden_node_subslot_utilization}.
	
	\section{Evaluation} \label{sec:evaluation} 
	The following subsections describe the evaluation of QMA in different scenarios. For this, it is compared to slotted and unslotted CSMA/CA, both of which are specified in IEEE 802.15.4 as multiple channel access schemes \cite{dsme_standard}. Evaluation is divided into three parts. At first, Sect.~\ref{sec:evaluation_hidden_node} describes QMA's capability to solve the hidden node problem without the additional overhead of  RTS / CTS messages. The second part in Sect.~\ref{sec:evaluation_iotlab} contains a validation of QMA in a physical testbed of the FIT IoT-LAB \cite{adjih_iotlab}. To reflect realistic use cases, experiments are conducted in a small tree topology and a dense star topology. Finally, Sect.~\ref{sec:evaluation_scalability} demonstrates QMA's scalability in large networks with an increasing number of nodes using the discrete event simulator OMNeT++. All results are presented with a 95\% confidence interval. For \QL we use $\lr=0.5$ and $\gamma=0.9$. 
	
	\subsection{Hidden-node problem}  
	\label{sec:evaluation_hidden_node}
\begin{figure}[h]
		\centering
		\includegraphics{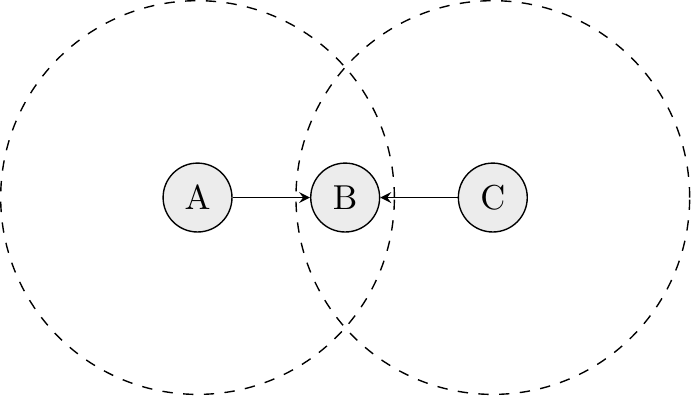}
		\caption{Hidden node problem with three nodes and the communication range of nodes \textit{A} and \textit{C}.}
		\label{fig:evaluation_hidden_node_topology}
	\end{figure}

    At first, a hidden node problem with three nodes, as depicted in Fig.~\ref{fig:evaluation_hidden_node_topology}, is presented. Here, a CCA at node \textit{A} or \textit{C} only fails if node \textit{B} is currently sending an ACK. Data packets sent by node \textit{A} or \textit{C} cannot be recognized by the opposite node using CCA. Therefore, the CCA is successful a large fraction of the time even though the channel is not idle. For this scenario, we consider primary traffic with a varying packet generation rate of $\delta$ packets per second. Nodes \textit{A} and \textit{C} generate 1000 data packets according to a Poisson distribution with mean $\lambda=\delta$ while node \textit{B} acts as the sink. 
    Generation of data packets starts after 100s to allow the MAC protocol to associate with the network and exchange management information. For comparison, experiments are also conducted with slotted and unslotted CSMA/CA. This allows for easy estimation of the PDR for different channel utilizations. The scenario is simulated in the discrete event simulator OMNeT++ with 15 repetitions per channel access scheme.    
	
	\subsubsection{Reliability} \label{sec:evaluation_hidden_node_reliability}
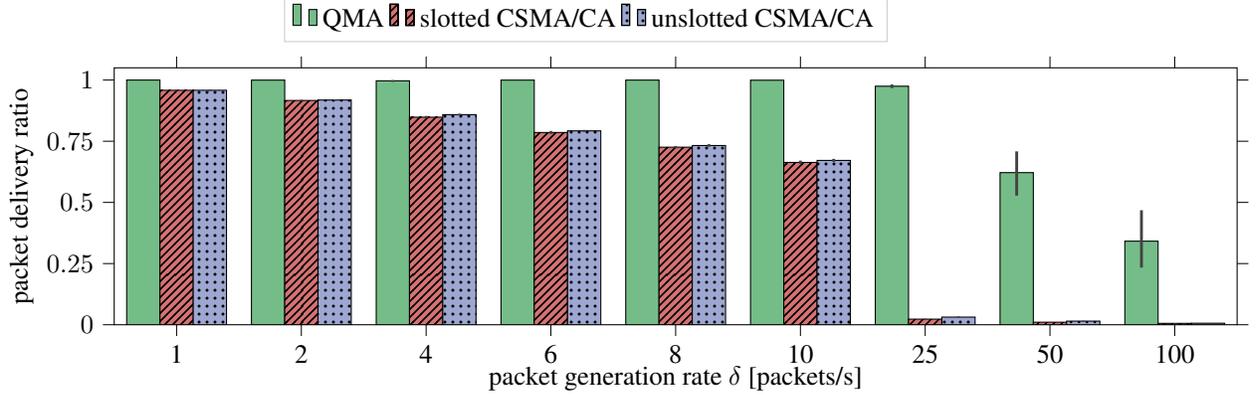
\begin{figure}[h]
		\centering
		\input{./plots/prr_hidden.tex}
		\caption{Packet delivery ratio of node \textit{A} and \textit{C} for different packet generation rates $\delta$.}
		\label{fig:evaluation_prr_hidden}
	\end{figure}

	Fig.~\ref{fig:evaluation_prr_hidden} shows the PDR of slotted and unslotted CSMA/CA and QMA for nodes \textit{A} and \textit{C} of the hidden node problem. As one can see, QMA outperforms both CSMA/CA mechanisms for all packet generation rates. At higher packet generation rates, e.g. $\delta=25$, QMA maintains a PDR of 97\% while both CSMA/CA mechanisms show a PDR below 3.5\%. However, the performance difference becomes smaller for lower rates due to two reasons. Firstly, the probability of a collision is smaller for fewer transmitted packets when using CSMA/CA. Secondly, QMA learns faster if more packets are transmitted because different subslots can be tested and only the best are used for transmission. If only a few packets have to be sent, QMA still works but the certainty about good and bad transmission subslots is not that high. 
	
\begin{figure}[h]
		\centering
		\input{./plots/queue_hidden.tex}
		\caption{Average queue level of nodes \textit{A} and \textit{C} for varying packet generation rates $\delta$.}
		\label{fig:evaluation_queue_hidden}
	\end{figure}
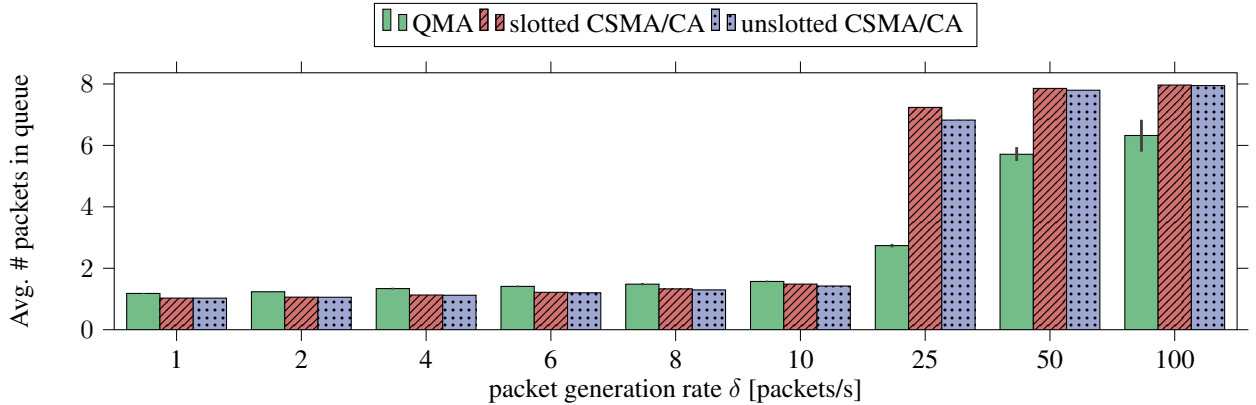

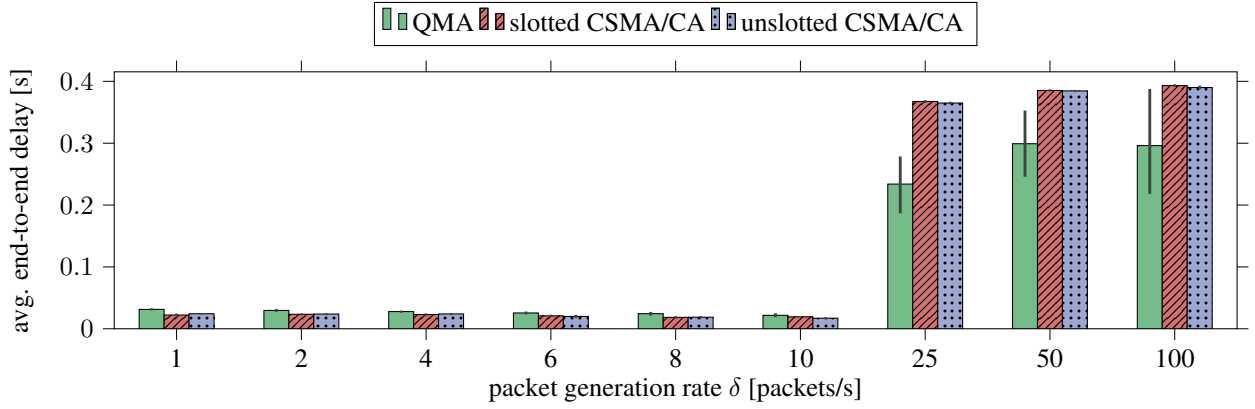
\begin{figure}[h]
		\centering
		\input{./plots/delay_hidden.tex}
		\caption{Average end-to-end delay of nodes \textit{A} and \textit{C} for varying packet generation rates $\delta$.}
		\label{fig:evaluation_delay_hidden}
	\end{figure}

	QMA's main idea is to hold back packets until an appropriate transmission subslot is reached. This can lead to increased queue levels, as illustrated in Fig.~\ref{fig:evaluation_queue_hidden}. For $\delta \le 10$ packets/s, QMA's queue level is slightly higher than the queue level of the CSMA/CA mechanisms, which consequently leads to a slightly increased end-to-end delay, as illustrated by Fig.~\ref{fig:evaluation_delay_hidden}, as packets remain in the queue longer. The end-to-end delay is the time between generation of a packet and its reception at the sink. The main cause of packet loss of the CSMA/CA mechanisms for $\delta \le 10$ packets/s are dropped packets due to too many retransmissions or, in rare cases, too many backoffs. For $\delta \ge 25$ packets/s, on the other hand, the probability for a collision is high, and the average queue level converges towards the maximum queue size of 8 packets. Here, most packets are lost due to queue drops as packets cannot be transmitted fast enough. QMA manages to reduce the queue level significantly for $\delta \ge 25$ packets/s because it finds hidden patterns in the traffic and holds back packets to avoid collisions with the other nodes. This also reduces delay, as packets stay in the queue for a shorter time, as shown in Fig.~\ref{fig:evaluation_queue_hidden} and Fig.~\ref{fig:evaluation_delay_hidden}.
	
	One should notice that the transmission behaviour of CSMA/CA is quite random. Sometimes many consecutive packets are transmitted successfully, sometimes a large number of transmissions fail due to the randomness in the exponential backoff algorithm. The same applies to QMA while it learns hidden traffic patterns, however, as soon as a feasible policy is found and if traffic conditions are stable, it offers deterministic reliability and delay. 
	
	\subsubsection{Convergence and adaptability} \label{sec:evaluation_hidden_node_convergence}
\begin{figure}[h]
		\centering
		\includegraphics{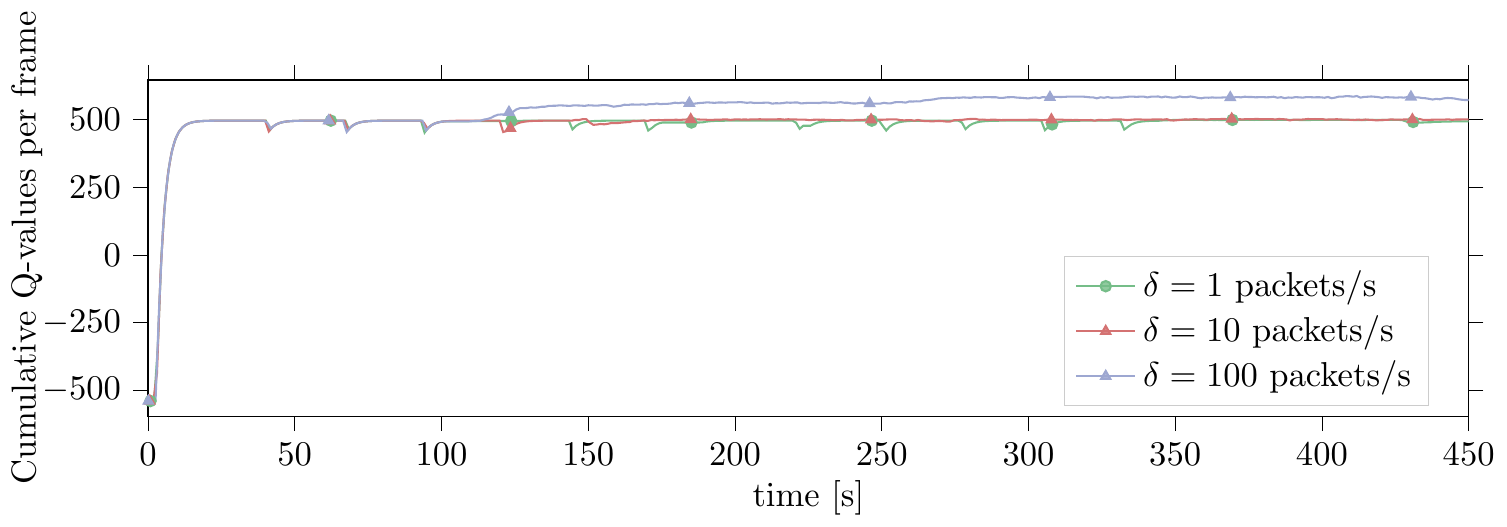}
		\caption{Cumulative Q-values per frame for different $\delta$.}
		\label{fig:evaluation_convergence_hidden}
	\end{figure}

	One of the main challenges in data-collection scenarios is adaptability to environmental changes, e.g., fluctuating traffic conditions and node failure. QMA offers good adaptability in these changing environments, as shown in Fig.~\ref{fig:evaluation_convergence_hidden}. It depicts the cumulative Q-values per frame over time, i.e., the sum of  Q-values for all subslots following the best policy at that time. This metric cannot express the quality of the learned solution but indicates its stability. In other words: A higher cumulative Q-value does not mean that an agent performs better than an agent with a lower cumulative Q-value. However, a stable cumulative Q-value indicates that the Q-table does not change and the policy is stable \cite{mnih2015human}. 
	
	Fig.~\ref{fig:evaluation_convergence_hidden} shows QMA's stability for different values of $\delta$. During the first 100 seconds, the MAC protocol exchanges management information via QMA. Afterwards, additional data packets are transmitted. It can be seen that QMA immediately reacts to the first transmitted management packets and settles on a stable policy about 9 seconds after the first transmission, as the rate of the management traffic is quite low. For $\delta=100$, the policy changes once more at 100 seconds when data packets start to be transmitted and after 265 seconds. The cumulative Q-value becomes constant about 30 seconds later. One should notice that the network is oversaturated for $\delta=100$ and nodes \textit{A} and \textit{C} compete for access to the channel. Thus agreeing on a common policy takes more time than in a non-saturated network. For $\delta=1$ and $\delta=10$, the policy does not change significantly when data packets start to be transmitted, as the management traffic exhibits about the same data rate.
	
\begin{figure}[h]
		\centering
		\includegraphics{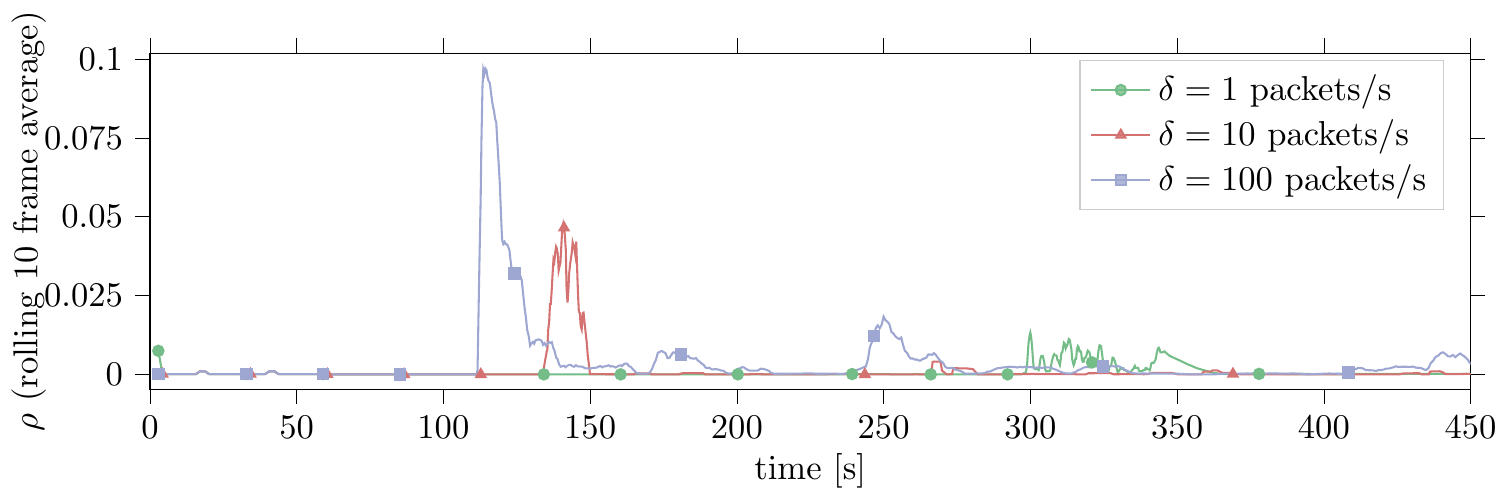}
		\caption{Exploration probability $\rho$ of parameter-based exploration over time using a rolling average over 10 frames for different values of $\delta$.}
		\label{fig:evaluation_eps_hidden}
	\end{figure}

	Fig.~\ref{fig:evaluation_eps_hidden} shows the according exploration probability $\rho$ using parameter-based exploration. As one can see, exploration is triggered earlier for higher $\delta$ because queues fill up more quickly. $\rho$ increases multiple times for $\delta=100$ as the network is oversaturated and nodes \textit{A} and \textit{C} compete for access to the shared medium. A maximum of $\rho=0.1$ is experienced for $\delta=100$, corresponding to a queue level of 6 packets. 
	
\begin{figure}[h]
		\centering
		\includegraphics{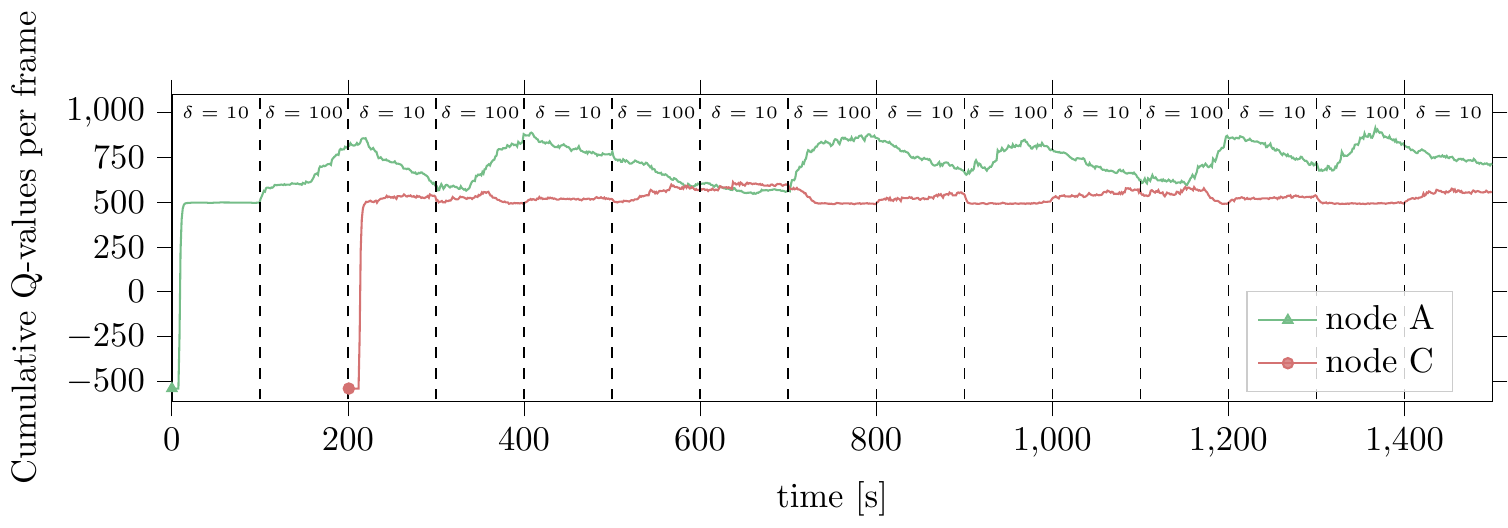}
		\caption{Cumulative Q-values per frame of nodes \textit{A} and \textit{C} for fluctuating traffic conditions.}
		\label{fig:evaluation_adaptability_hidden}
	\end{figure}

	To illustrate QMA's adaptability, the scenario is slightly altered  to incorporate fluctuating traffic conditions. Node \textit{A} alternately generates packets with $\delta=10$ and $\delta=100$ for 100 seconds respectively. Node \textit{C} generates a constant $\delta=25$ packets/s but joins the network 100 seconds after node \textit{A}. The positions of the nodes remain as shown in Fig.~\ref{fig:evaluation_hidden_node_topology}. Fig.~\ref{fig:evaluation_adaptability_hidden} shows the cumulative Q-values per frame. The dashed lines indicate when the traffic pattern of node \textit{A} changes. 
	
	As one can see, node \textit{A} immediately reacts to changes in its packet generation pattern with increasing and decreasing Q-values. The network is oversaturated for $\delta=100$ and not saturated for $\delta=10$. Thus, node \textit{C} tries to acquire more transmission slots when node \textit{A} generates packets with $\delta=10$. This is indicated by the increase of the cumulative Q-value of node \textit{C} for $\delta=10$ at node \textit{A}. The change of Q-values is triggered by parameter-based exploration, as packets start to pile up in the queue of node \textit{C} when node \textit{A} transmits with $\delta=100$. Thus, QMA's reaction time is mainly dictated by the exploration rate assigned to different queue levels, as described in Sect.~\ref{sec:parameter_based_exploration}, and can be increased and decreased according to the given scenario. At last, joining the network late does not influence the performance of node \textit{C}, since a stable policy is still found. 
	
	\subsubsection{Subslot utilization} \label{sec:evaluation_hidden_node_subslot_utilization}
	Fig.~\ref{fig:slot_utilization_hidden_1} to \ref{fig:slot_utilization_hidden_100} show the utilization of subslots by node \textit{A} and \textit{C} for different $\delta$. Fig.~\ref{fig:evaluation_eps_hidden} illustrates that exploration is triggered at different times depending on $\delta$. Thus, Fig.~\ref{fig:slot_utilization_hidden_1} to \ref{fig:slot_utilization_hidden_100} depict the subslot utilization after the first exploration phase, i.e., at 170 seconds for $\delta=100$, 150 seconds for $\delta=10$, and 370 seconds for $\delta=1$. Additionally, the final policy is shown. If no action is shown, \QBACKOFF is executed.
	
	As one can see, a collision-free schedule of subslots is created for all values of $\delta$, i.e., nodes \textit{A} and \textit{C} never select action \QCCA or \QSEND in the same subslot. Additionally, there are no \QSEND actions in adjacent subslots. This is important because transmissions span over up to 3 subslots. In Fig.~\ref{fig:slot_utilization_hidden_1a} node \textit{A} chooses action \QSEND, however, for the hidden node problem, there is almost no difference between \QSEND and \QCCA as a CCA is almost always successful. The only exception is when an ACK by node \textit{B} is received by chance. Thus, node \textit{A} selects an additional transmission slot in the final policy to maximize the total expected reward. 
	
	Fig.~\ref{fig:slot_utilization_hidden_10} and \ref{fig:slot_utilization_hidden_100} show that the first exploration phase results in many transmission slots at a single node. Additionally, Fig.~\ref{fig:slot_utilization_hidden_10} shows that node \textit{A} and \textit{C} divide the medium among themselves in the final policy. Such patterns arise when the exploration probability is high, i.e., when the queue is full using parameter-based exploration. In this case, many consecutive random actions are executed as shown in Fig.~\ref{fig:slot_utilization_hidden_100a}. If no more packets are available for transmission, no action is selected. In the next iteration, a node uses the newly allocated transmissions subslots, potentially leaving later subslots unused because the queue is empty. This allows the other node to allocate transmission slots in the later subslots and, eventually, node \textit{A} gives up later transmission subslots because it overhears ACK packets for node \textit{C}. In this scenario, almost all subslots are utilized. 
	
	\begin{figure}[h]
		\centering
		\begin{subfigure}[b]{0.49\textwidth}
			\centering
			\input{./plots/slot_utilization_hidden_1a.tex}
			\caption{Slot utilization after 370 seconds.}
			\label{fig:slot_utilization_hidden_1a}
		\end{subfigure} 
		\hfill
		\begin{subfigure}[b]{0.49\textwidth}
			\centering
			\input{./plots/slot_utilization_hidden_1b.tex}
			\caption{Slot utilization for final policy.}
			\label{fig:slot_utilization_hidden_1b}
		\end{subfigure}
		\caption{Slot utilization for $\delta=1$ after 370 seconds and for the final policy.}
		\label{fig:slot_utilization_hidden_1}
	\end{figure}
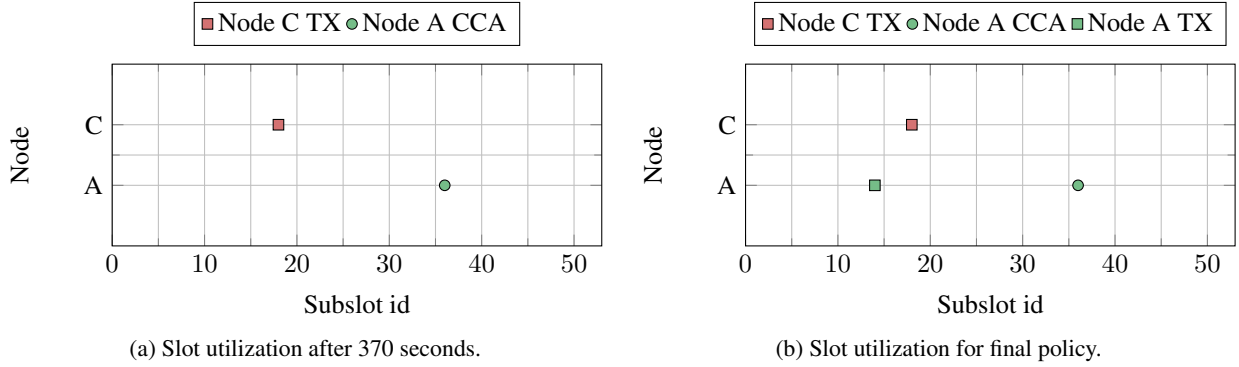
	
	\begin{figure}[h]
		\centering
		\begin{subfigure}[b]{0.49\textwidth}
			\centering
			\input{./plots/slot_utilization_hidden_10a.tex}
			\caption{Slot utilization after 150 seconds.}
			\label{fig:slot_utilization_hidden_10a}
		\end{subfigure} 
		\hfill
		\begin{subfigure}[b]{0.49\textwidth}
			\centering
			\input{./plots/slot_utilization_hidden_10b.tex}
			\caption{Slot utilization for final policy.}
			\label{fig:slot_utilization_hidden_10b}
		\end{subfigure}
		\caption{Slot utilization for $\delta=10$ after 150 seconds and for the final policy.}
		\label{fig:slot_utilization_hidden_10}
	\end{figure}
	
	\begin{figure}[h]
		\centering
		\begin{subfigure}[b]{0.49\textwidth}
			\centering
			\input{./plots/slot_utilization_hidden_100a.tex}
			\caption{Slot utilization after 170 seconds.}
			\label{fig:slot_utilization_hidden_100a}
		\end{subfigure} 
		\hfill
		\begin{subfigure}[b]{0.49\textwidth}
			\centering
			\input{./plots/slot_utilization_hidden_100b.tex}
			\caption{Slot utilization for final policy.}
			\label{fig:slot_utilization_hidden_100b}
		\end{subfigure}
		\caption{Slot utilization for $\delta=100$ after 170 seconds and for the final policy.}
		\label{fig:slot_utilization_hidden_100}
	\end{figure}
	
	In contrast to that, many subslots are not utilized for $\delta=1$ and $\delta=10$, i.e., action \QBACKOFF is chosen. An obvious implication is that there is still a lot of room for scheduling additional transmissions. However, one should be aware that transmissions can last for more than one subslot and thus there is a necessity for those idle slots. 
	
	\subsection{Verification in IoT-Testbed} \label{sec:evaluation_iotlab}
\begin{figure}[h]
		\centering
		\begin{minipage}[b]{0.4\textwidth}
			\centering
			\includegraphics{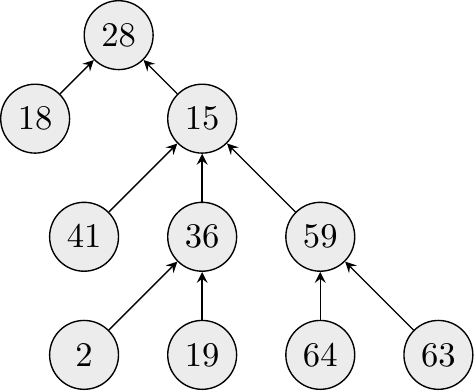}
			\caption{Logical routing tree for evaluation in the Strasbourg testbed of the FIT IoT-Lab.}
			\label{fig:iotlab_topology}
		\end{minipage}
		\hspace{1cm}
		\begin{minipage}[b]{.45\textwidth}
			\centering
			\includegraphics{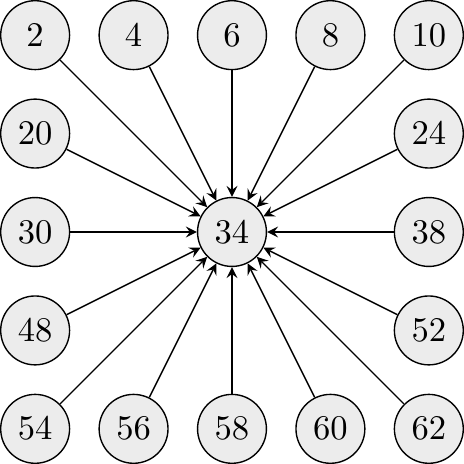}
			\caption{Physical node positions for a star topology in the Strasbourg testbed of the FIT IoT-Lab.}
			\label{fig:iotlab_topology_star}
		\end{minipage}
	\end{figure}

	To verify QMA's resource-efficiency and apply it in a scenario with physical nodes, some experiments are conducted in the \textit{Strasbourg} testbed of the FIT IoT-LAB \cite{adjih_iotlab}. For this, CometOS, a component-based operating system for wireless networks \cite{unterschutz2012cross}, is utilized to run openDSME on M3 Open Node boards featuring 32-bit Cortex M3 CPUs with a maximum clock speed of 72 MHz, 64 KB of RAM, and 256 KB of ROM \footnote{\url{https://iot-lab.github.io/docs/boards/iot-lab-m3/}}. The Cortex M3 does not feature a floating-point unit, thus an efficient solution is required. Two scenarios are evaluated, the fist of which is a tree topology with 10 nodes and depth 4 as illustrated in Fig.~\ref{fig:iotlab_topology} and the second is a star topology with 17 nodes as shown in Fig.~\ref{fig:iotlab_topology_star}. The tree topology is generated using an algorithm proposed by Kauer et al. \cite{kauer2018constructing}. According to their description, transmission power is set to -9 dBm and sensitivity is set to -72 dBm. For the star topology, a transmission power of 3 dBm and a sensitivity of -90 dBm is used.
	
    The setup is similar to Sect.~\ref{sec:evaluation_hidden_node}, i.e, 1000 data packets are generated according to a Poisson distribution and 
    experiments are repeated 10 times for each channel access scheme. Simulations in Sect.~\ref{sec:evaluation_hidden_node_reliability} and experiments on hardware have shown that slotted and unslotted CSMA/CA perform almost the same. Thus, the following sections feature only a comparison of unslotted CSMA/CA and QMA.

	\subsubsection{Reliability}
\begin{figure}[h]
		\centering
		\input{./plots/iotlab_star_prr.tex}
		\caption{Packet delivery ratio for different nodes of a star topology in the FIT IoT-Lab with $\delta=10$.}
		\label{fig:evaluation_iotlab_star_prr}
	\end{figure}
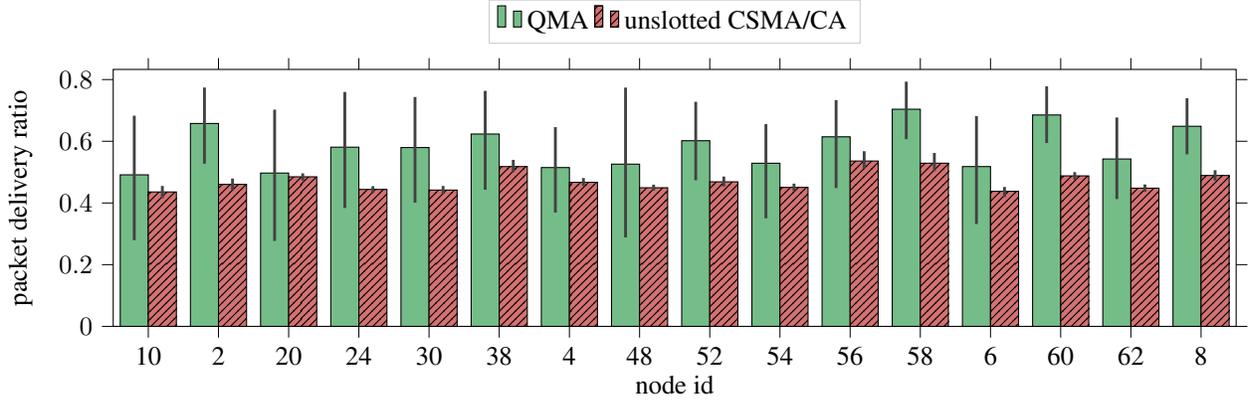

	In this section, QMA is verified on hardware in the FIT IoT-LAB. For this, Fig.~\ref{fig:evaluation_iotlab_tree_prr} and Fig.\ref{fig:evaluation_iotlab_star_prr} show the PDR of the tree topology and the star topology, respectively. Both scenarios are evaluated with $\delta=10$. As depicted in Fig.~\ref{fig:evaluation_iotlab_tree_prr}, QMA achieves a higher PDR at all nodes of the tree topology than unslotted CSMA/CA. This supports the results from Sect.~\ref{sec:evaluation_hidden_node} because the tree topology exhibits several hidden node problems. Only transmissions of parents and children and siblings in the tree interfere with each other. The same applies to the results of the star topology, as shown in Fig.~\ref{fig:iotlab_topology_star}. Here, all nodes can hear each other so that the PDR of QMA and unslotted CSMA/CA is closer than in the tree scenario as CSMA/CA's CCA prevents many collisions. 
	
	\begin{figure}[h]
		\centering
		\input{./plots/iotlab_tree_prr.tex}
		\caption{Packet delivery ratio for different nodes of a tree topology in the FIT IoT-Lab with $\delta=10$.}
		\label{fig:evaluation_iotlab_tree_prr}
	\end{figure}
	
	An interesting observation is that nodes achieve a fair distribution of transmission subslots using QMA, even though no fairness constraint is explicitly handled in its reward function. For example, it could have happened that a small number of nodes acquire all available subslots for action \QSEND while the other nodes starve. Fairness arises due to two reasons. Firstly, QMA utilizes a cooperative Q-learning approach, as discussed in Sect.~\ref{sec:self_synchronization}. This helps to build a schedule but does not completely ensure fairness. However, \PAREX results in many random actions if a node possesses too few transmission subslots. Thus, if a node used all subslots for \QSEND, its transmissions would be interfered by these random actions and it gives up the transmission slot. Another node can now acquire this subslots for \QSEND. This process balances itself, given that the network is not too oversaturated. 
	
	At last, energy measurements in the IoT-LAB show no difference between QMA and unslotted CSMA/CA in terms of power consumption. Thus, one can conclude that QMA increases reliability without introducing an energy-overhead. In fact, both multiple access schemes conduct about the same number of transmission attempts, however, using CSMA/CA many failed CCAs and retransmissions occur. 
	
	
	\subsection{Scalability in data-collection scenarios} \label{sec:evaluation_scalability}
\begin{figure}[h]
		\centering
		\includegraphics{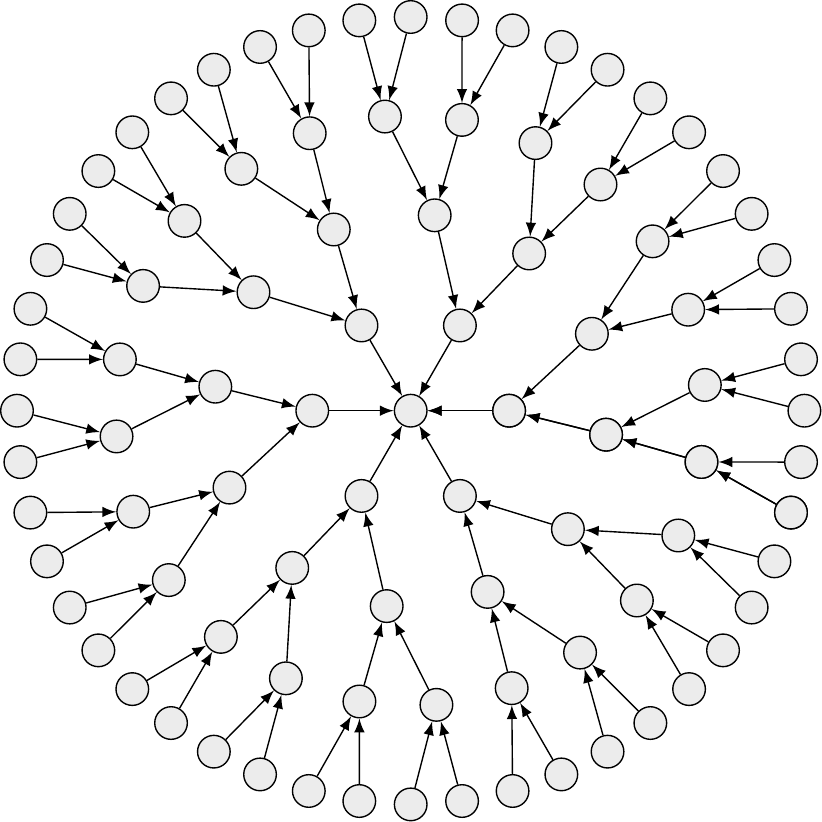}
		\caption{Concentric topology with 91 nodes and according routing tree.}
		\label{fig:concentric_topology}
	\end{figure}

	At last, QMA's scalability is investigated in a concentric topology with an increasing number of nodes, as shown in Fig.~\ref{fig:concentric_topology}. A number of 7, 19, 43, and 91 nodes is evaluated, corresponding to 1 to 4 rings around the center node.  Experiments are conducted with OMNeT++. In contrast to Sect.~\ref{sec:evaluation_hidden_node}~and~\ref{sec:evaluation_iotlab}, QMA is evaluated in a realistic scenario for secondary traffic during the CAP of IEEE 802.15.4 DSME, i.e., for (de)allocation of GTS and transmission of broadcasts. IEEE 802.15.4 and extensions building upon it, e.g. ZigBee and WirelessHart, gained a lot of popularity for industrial application over the last few years and, thus, seems to be a good fit for the evaluation \cite{jecan2018dual}. The latter are generated by the routing protocol, \textit{greedy perimeter stateless routing} (GPSR), for route discovery. The center node acts as a sink and every other node generates data packet according to a Poisson distribution with fluctuating rate and sends them towards the sink. More specifically, nodes alternately generate data packets with $\delta=1$ packet/s and $\delta=10$ packets/s for 5 seconds respectively. All data packets are transmitted during the CFP but cause secondary traffic for GTS (de)allocation. A single packet is transmitted per GTS and experiments are repeated 15 times with 1000 evaluation packets and a warmup time of 200 seconds to allow for network formation. It should be noted that such data-collection scenarios often suffer from multiple hidden node problems \cite{Telematik_Kauer_2019_Diss}. 
	
	\subsubsection{Reliability}
\begin{figure}[h]
		\centering
		\input{plots/cap_prr_scalability.tex}
		\caption{PDR for secondary traffic during the CAP of IEEE 802.15.4 DSME for different numbers of nodes.}
		\label{fig:evaluation_scalability_prr}
	\end{figure}
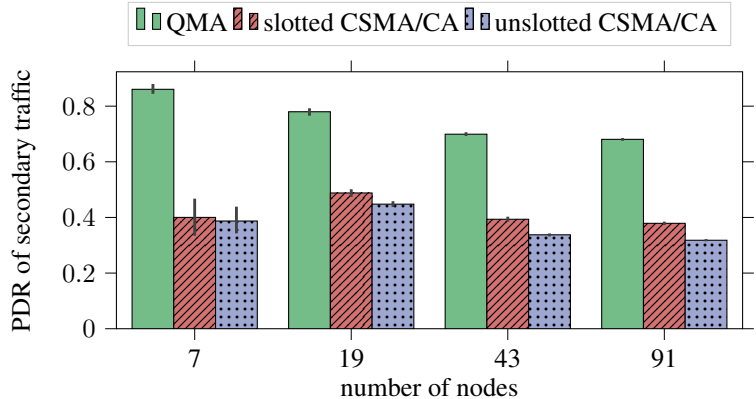

	Fig.~\ref{fig:evaluation_scalability_prr} shows the PDR of DSME's secondary traffic for an increasing number of nodes. As one can see, QMA outperforms slotted and unslotted CSMA/CA for any number of nodes, however, the performance difference is larger for a smaller number of nodes. That is because QMA can learn a collision-free policy faster if fewer nodes are within interference range because the probability of randomly selecting an unused subslots is higher. In comparison to QMA and as shown in Sect.~\ref{sec:evaluation_hidden_node}, CSMA/CA cannot fully solve the hidden node problem. Thus its PDR is significantly lower for all numbers of nodes. Here, the number of nodes does not significantly influence the performance of CSMA/CA since they are placed far enough from each other. QMA's increased PDR during the CAP also results in an increased PDR of primary traffic of about 10\% for any number of nodes, as GTS are (de)allocated faster and thus fewer packets are dropped due to a full queue. 
	
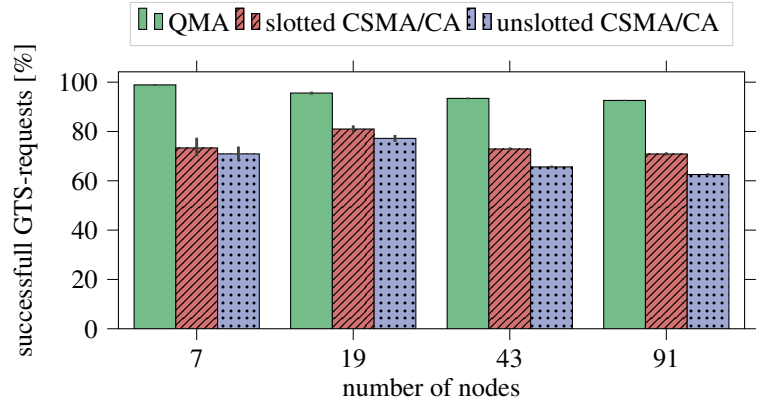
\begin{figure}[h]
		\centering
		\input{plots/cap_handshake_req.tex}
		\caption{Percentage of successfully transmitted \GTSREQUEST for different numbers of nodes.}
		\label{fig:evaluation_scalability_handshake_req}
	\end{figure}

	Fig.~\ref{fig:evaluation_scalability_handshake_req} demonstrates QMA's capability of learning the hidden patterns during slot allocations. As shown in Sect.~\ref{sec:theoretical_considerations}, slot (de)allocations through 3-way handshakes require a high PDR during the CAP. The increased PDR during the CAP significantly increases the number of successfully transmitted \GTSREQUEST messages, which are used to initialize the 3-way handshake. QMA's advantage seems to diminish for \GTSRESPONSE and \GTSNOTIFY messages. For both message types, QMA manages to transmit 99\% messages successfully, while CSMA/CA achieves about 93\% and 99\% messages of \GTSRESPONSE and \GTSNOTIFY messages successfully, respectively. However, in absolute numbers, CSMA/CA only transmits 2/3 of the \GTSNOTIFY messages that QMA transmits so that both methods result in about 1\% failed \GTSNOTIFY messages.
	
	\section{Conclusion}
	In this work, we propose \textit{QMA}, a Q-learning based multiple access scheme for wireless sensor networks. Whilst many current MAC-protocols for the Industrial Internet of Things feature TDMA-based channel access for primary traffic to increase reliability and scalability, they still require contention-based channel access for secondary traffic, e.g., for broadcast transmission and slot allocation. Especially in networks with fluctuating primary traffic, where TDMA-slots are constantly allocated and deallocated, most contention-based protocols do not provide sufficient reliability. QMA's core idea is to learn subtle but hidden patterns in the primary and secondary traffic which are not utilized by present contention-base channel access schemes. This way, QMA can adjust transmission slots so that no collisions occur. Thereby, QMA is designed to be resource-efficient to allow execution on small embedded devices. 
	
	We demonstrate QMA's effectiveness in three different scenarios. At first, it is utilized for primary traffic in a hidden node problem with three nodes and varying packet generation rates. Simulations show that it is possible to increase the number of transmitted packets from 10 packets/s to 50 packets/s in comparison to CSMA/CA while maintaining the same packet delivery ratio. In the second scenario, a real implementation of QMA is deployed and verified in a FIT IoT-LAB testbed. QMA and CSMA/CA consume the same amount of energy for all configurations because the number of transmission attempts is the same. However, for CSMA/CA these attempts often result in backoffs due to failed CCA, while QMA manages to transmit the packet successfully. Additionally, we show that QMA achieves fairness and increases reliability without consuming more energy. At last, QMA's scalability is evaluated for secondary traffic in IEEE 802.15.4 DSME in a realistic scenario with an increasing number of nodes. QMA achieves a higher packet delivery ratio than CSMA/CA and manages to (de)allocate up to twice more TDMA-slots per second. 
	
	In a future work, we will investigate additional techniques to further reduce QMA's memory overhead. For example, by carefully choosing rewards, it should be possible to significantly reduce the entries of the Q-table so that only 2 - 8 Bit are required and floating-point operations are avoided. 
	
	All in all, we believe that QMA is a practical realization of a more reliable contention-based channel access for the IIoT. The benefits of QMA are an indicator that machine learning techniques bear a high potential for communication protocols. 
	
	\bibliography{document}{}
	\bibliographystyle{plain}
	
	\newpage
	\appendix
	
	\section{DSME in a Nutshell} \label{sec:dsme}	
\begin{figure}[h]
		\centering
		\includegraphics[]{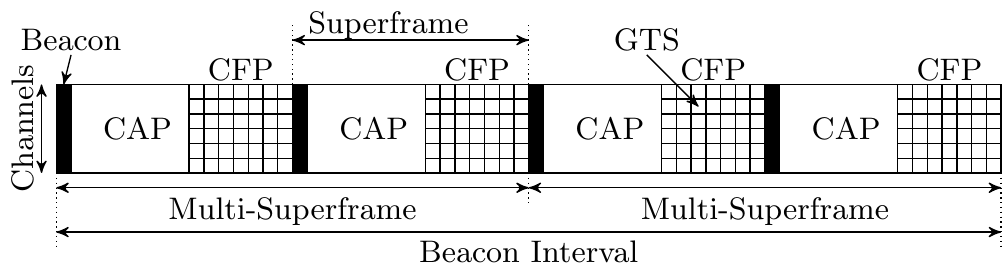}
		\caption{Structure of a DSME multi-superframe.}
		\label{fig:dsme_multi_superframe}
	\end{figure}

	IEEE 802.15.4 DSME divides time into multi-superframes and further into superframes as shown in Fig.~\ref{fig:dsme_multi_superframe}. Superframes consist of 16 time slots, the first of which is used to transmit beacons containing network and time information. The following 15 slots are separated into two phases: a \textit{contention access period} (CAP) with 8 time slots and a \textit{contention free period} (CFP) with 7 time slots. The CFP is further subdivided into \textit{guaranteed time slots} (GTS) which are spread over time and frequency and allow exclusive access to the shared medium. Before usage, they must be allocated during the CAP through a 3-way-handshake. Additionally, the CAP can be used for exchanging management messages and broadcasts via CSMA/CA. 
	
\begin{figure}[h]
		\centering
		\includegraphics[]{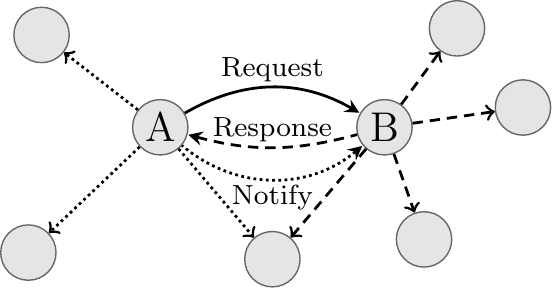}
		\caption{3-way slot allocation handshake in IEEE 802.15.4 DSME.}
		\label{fig:3_way_handshake}
	\end{figure}

	Fig.~\ref{fig:3_way_handshake} shows the 3-way GTS allocation handshake of IEEE 802.15.4 DSME. It is initialized by the unicast transmission of a GTS-request from a node A to a node B. B responds with a broadcast of the GTS-response to inform all nodes in its neighbourhood about the GTS that is going to be allocated. When A receives the GTS-response, it finalizes the handshake with a broadcasted GTS-notify to also inform all nodes in its neighbourhood about the GTS that is allocated. If any of A's or B's neighbours have already allocated the GTS, i.e., a duplicate allocation occurs, or if any of the 3 messages is lost, the GTS allocation is rolled back using the same 3-way handshake. Afterwards, A and B can try to allocate another GTS.  
	
	\subsection{Case study: 3-way handshake in DSME} 
	\label{sec:theoretical_considerations}
\begin{figure}[h]
		\centering
		\includegraphics{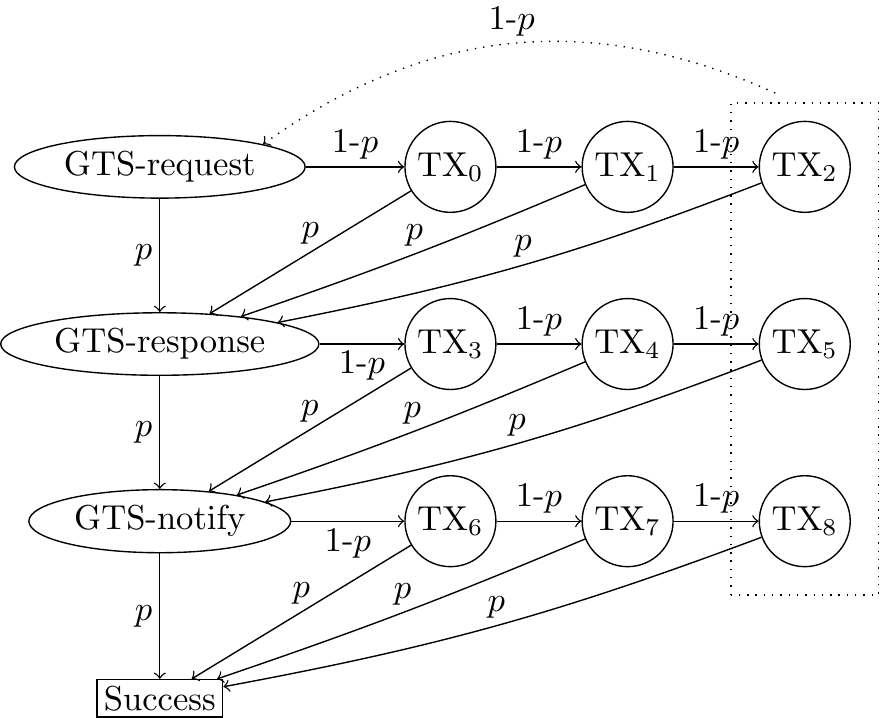}
		\caption{Absorbing markov chain of the GTS-negotiation where $p$ is the probability of a successful packet transmission.} 
		\label{fig:markov_handshake}
	\end{figure}

	To illustrate the importance of a high probability for a successful packet transmission during the CAP, one can calculate the average number of required packets to allocate a single GTS. Especially in scenarios with highly fluctuating traffic, it is necessary to keep the number of sent packets close to the minimum of 3 (3-way-handshake: \GTSREQUEST, \GTSRESPONSE, \GTSNOTIFY) so that as many GTS as possible can be successfully allocated within a single CAP. An absorbing Markov chain of the handshake is shown in Fig.~\ref{fig:markov_handshake}. The \textit{TX}-states symbolize CSMA/CA retransmissions of the respective handshake message, which is dropped after 3 retries.
	
	The absorbing Markov chain from 
	Fig.~\ref{fig:markov_handshake} with $t=12$ transient states 
	and $r=1$ absorbing states can be represented in its 
	canonical form with transition matrix P as  
	\begin{align}
		P = \begin{pmatrix}
			Q & R \\
			\mathbf{0}  & I_r
		\end{pmatrix},
	\end{align}
	where $Q$ is a $t \times t$ matrix, describing the 
	probability 
	to transition from a transient state to another transient 
	state, $R$ is a $t 
	\times r$ matrix, describing the probability to transition 
	from a transient state to an absorbing state, $\mathbf{0}$ 
	is an $r 
	\times t$ zero matrix and $I_r$ is an $r \times r$ identity 
	matrix. For above markov chain $Q$ is defined as 
	\begin{align}
		\footnotesize
		Q = \begin{pmatrix}
			0 & p & 0 & 1-p & 0 & 0 & 0 & 0 & 0 & 0 & 0 & 0 \\
			0 & 0 & p & 0 & 0 & 0 & 1-p & 0 & 0 & 0 & 0 & 0 \\
			0 & 0 & 0 & 0 & 0 & 0 & 0 & 0 & 0 & 1-p & 0 & 0 \\
			0 & p & 0 & 0 & 1-p & 0 & 0 & 0 & 0 & 0 & 0 & 0 \\
			0 & p & 0 & 0 & 0 & 1-p & 0 & 0 & 0 & 0 & 0 & 0 \\
			1-p & p & 0 & 0 & 0 & 0 & 0 & 0 & 0 & 0 & 0 & 0 \\
			0 & 0 & p & 0 & 0 & 0 & 0 & 1-p & 0 & 0 & 0 & 0 \\
			0 & 0 & p & 0 & 0 & 0 & 0 & 0 & 1-p & 0 & 0 & 0 \\
			1-p & 0 & p & 0 & 0 & 0 & 0 & 0 & 0 & 0 & 0 & 0 \\
			0 & 0 & 0 & 0 & 0 & 0 & 0 & 0 & 0 & 0 & 1-p & 0 \\
			0 & 0 & 0 & 0 & 0 & 0 & 0 & 0 & 0 & 0 & 0 & 1-p \\
			1-p & 0 & 0 & 0 & 0 & 0 & 0 & 0 &0 & 0 & 0 & 0 \\
		\end{pmatrix}.
	\end{align}
	This allows us to calculate the fundamental matrix $N$ as 
	\begin{align}
		N = (I_t - Q)^{-1},
	\end{align}
	where $I_t$ is a $t \times t$ identity matrix, and the 
	expected number of steps $S$ until absorption as 
	\begin{align}
		S = N\mathbf{1},
	\end{align}
	where $\mathbf{1}$ is a vector of size $t$ with all ones. 
	
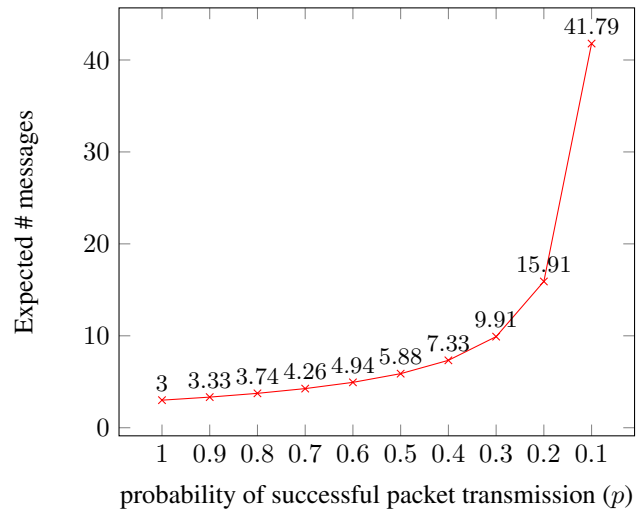
\begin{figure}[h]
		\centering
		\begin{tikzpicture}
			\begin{axis} [
				xlabel = probability of successful packet transmission ($p$), 
				xtick = data,
				ylabel = Expected \# messages, 
				nodes near coords,
				every node near coord/.append 
				style={color=black, 
					font=\footnotesize},
				x dir=reverse,
				]
				\addplot[color=red,mark=x] coordinates {
					(0.1, 41.78822041)
					(0.2, 15.90883469)
					(0.3, 9.90987542)
					(0.4, 7.33224265)
					(0.5, 5.88333333)
					(0.6, 4.93712096)
					(0.7, 4.25880878)
					(0.8, 3.74200321)
					(0.9, 3.33233334)
					(1.0, 3)
				};
			\end{axis}
		\end{tikzpicture}
		\caption{Expected number of sent messages during the 
			3-way GTS handshake for different values of $p$.}
		\label{fig:markov_handshake_prr}
	\end{figure}

	Fig.~\ref{fig:markov_handshake_prr} shows the expected number of steps until absorption, i.e., the number of required messages until a 3-way handshake is completed successfully and a GTS is allocated, for different values of $p$. As one can see, the number of required messages increases exponentially for a decreasing $p$. This 
	illustrates the necessity for a high probability of a successful transmission during the CAP to ensure timely (de)allocation of GTS. 
	
\end{document}

%% file: plots/parameter_based_exploration.tex
\pgfplotsset{compat = 1.3}
\usetikzlibrary{pgfplots.groupplots}

\begin{tikzpicture}
    \begin{groupplot} [
    height=6cm,
    group style={
    	group name=my fancy plots,
    	group size=2 by 1,
    	xticklabels at=edge bottom,
    	horizontal sep=0pt,
    },
    ylabel = Random action probability $\rho$, 
    ymin=-0.07, ymax=0.35,
    nodes near coords,
    every node near coord/.append style={color=black, 
    	font=\footnotesize,
    	yshift=-0.55cm,
    	xshift=0.1cm,
    	/pgf/number format/fixed,
    	/pgf/number format/precision=4
    },
    ]
    \nextgroupplot[xmin=-9,xmax=-6.5,
    	xtick={-8, -7},
    	axis y line=left,
    	width=4cm,
    	axis line style={-},
    ]
    \addplot[color=red,mark=x] coordinates {
    	(-8,0)
    	(-7,0)
    	(1,0)
    };
    \nextgroupplot[xmin=-0.8,xmax=9,
    	xtick={0,1,2,3,4,5,6,7,8},
    	axis y line=right,
    	axis x discontinuity=parallel,
    	xlabel = local queue level - neighbors' avg. queue level,
    	xlabel style={xshift=-1.25cm},
    	width=11cm,
    	yticklabels={},
    	axis line style={-},
    	ylabel={}
    ]
    \addplot[color=red,mark=x] coordinates {
    	(-3,0)
        (0, 0)
        (1, 0.0001)
        (2, 0.001)
        (3, 0.008)
        (4, 0.02)
        (5, 0.05)
        (6, 0.1)
        (7, 0.18)
        (8, 0.3)
    };
    \end{groupplot}
\end{tikzpicture}

%% file: plots/prr_hidden.tex
\begin{tikzpicture}

\begin{axis}[
width=\FIGWIDTH,
height=\FIGHEIGHT,
legend columns=3, 
legend style={fill opacity=1.8, draw opacity=1, text opacity=1, anchor=south, draw=white!80.0!black, at={(0.42, 1.1)}},
tick align=outside,
tick pos=both,
x grid style={white!69.01960784313725!black},
xlabel={packet generation rate $\delta$ [packets/s]},
xmin=-0.5, xmax=8.5,
xtick style={draw=black,color=black},
y grid style={white!69.01960784313725!black},
ylabel={packet delivery ratio},
ymin=0, ymax=1.05,
ytick style={draw=black,color=black},
xticklabels={100, 50, 25, 10, 8, 6, 4, 2, 1},
ytick={0,0.25,0.5,0.75,1},
x dir=reverse,
xtick={0,1,2,3,4,5,6,7,8,9},
xlabel style={yshift=0.1cm}
]
\draw[draw=black,fill=color0,draw opacity=1] (axis cs:0.133333333333333,0) rectangle (axis cs:0.4,0.3416);
\addlegendimage{ybar,ybar legend,draw=black,fill=color0,draw opacity=1};
\addlegendentry{QMA}

\draw[draw=black,fill=color0,draw opacity=1] (axis cs:1.13333333333333,0) rectangle (axis cs:1.4,0.6217);
\draw[draw=black,fill=color0,draw opacity=1] (axis cs:2.13333333333333,0) rectangle (axis cs:2.4,0.9748);
\draw[draw=black,fill=color0,draw opacity=1] (axis cs:3.13333333333333,0) rectangle (axis cs:3.4,0.999366666666667);
\draw[draw=black,fill=color0,draw opacity=1] (axis cs:4.13333333333333,0) rectangle (axis cs:4.4,0.999766666666667);
\draw[draw=black,fill=color0,draw opacity=1] (axis cs:5.13333333333333,0) rectangle (axis cs:5.4,0.9997);
\draw[draw=black,fill=color0,draw opacity=1] (axis cs:6.13333333333333,0) rectangle (axis cs:6.4,0.9968);
\draw[draw=black,fill=color0,draw opacity=1] (axis cs:7.13333333333333,0) rectangle (axis cs:7.4,0.9997);
\draw[draw=black,fill=color0,draw opacity=1] (axis cs:8.13333333333333,0) rectangle (axis cs:8.4,0.999966666666667);
\draw[draw=black,fill=color1, postaction={pattern=north east lines}, postaction={pattern=north east lines}, postaction={pattern=north east lines},draw opacity=1] (axis cs:-0.133333333333333,0) rectangle (axis cs:0.133333333333333,0.00576666666666667);
\addlegendimage{ybar,ybar legend,draw=black,fill=color1, postaction={pattern=north east lines}, postaction={pattern=north east lines}, postaction={pattern=north east lines},draw opacity=1};
\addlegendentry{slotted CSMA/CA}

\draw[draw=black,fill=color1, postaction={pattern=north east lines}, postaction={pattern=north east lines}, postaction={pattern=north east lines},draw opacity=1] (axis cs:0.866666666666667,0) rectangle (axis cs:1.13333333333333,0.0104333333333333);
\draw[draw=black,fill=color1, postaction={pattern=north east lines}, postaction={pattern=north east lines}, postaction={pattern=north east lines},draw opacity=1] (axis cs:1.86666666666667,0) rectangle (axis cs:2.13333333333333,0.0229666666666667);
\draw[draw=black,fill=color1, postaction={pattern=north east lines}, postaction={pattern=north east lines}, postaction={pattern=north east lines},draw opacity=1] (axis cs:2.86666666666667,0) rectangle (axis cs:3.13333333333333,0.6634);
\draw[draw=black,fill=color1, postaction={pattern=north east lines}, postaction={pattern=north east lines}, postaction={pattern=north east lines},draw opacity=1] (axis cs:3.86666666666667,0) rectangle (axis cs:4.13333333333333,0.725666666666667);
\draw[draw=black,fill=color1, postaction={pattern=north east lines}, postaction={pattern=north east lines}, postaction={pattern=north east lines},draw opacity=1] (axis cs:4.86666666666667,0) rectangle (axis cs:5.13333333333333,0.7857);
\draw[draw=black,fill=color1, postaction={pattern=north east lines}, postaction={pattern=north east lines}, postaction={pattern=north east lines},draw opacity=1] (axis cs:5.86666666666667,0) rectangle (axis cs:6.13333333333333,0.848233333333333);
\draw[draw=black,fill=color1, postaction={pattern=north east lines}, postaction={pattern=north east lines}, postaction={pattern=north east lines},draw opacity=1] (axis cs:6.86666666666667,0) rectangle (axis cs:7.13333333333333,0.916533333333333);
\draw[draw=black,fill=color1, postaction={pattern=north east lines}, postaction={pattern=north east lines}, postaction={pattern=north east lines},draw opacity=1] (axis cs:7.86666666666667,0) rectangle (axis cs:8.13333333333333,0.958166666666667);

\draw[draw=black,fill=color2, postaction={pattern=dots},draw opacity=1] (axis cs:-0.4,0) rectangle (axis cs:-0.1333333333333,0.00626666666666667);
\addlegendimage{ybar,ybar legend,draw=black,fill=color2, postaction={pattern=dots},draw opacity=1};
\addlegendentry{unslotted CSMA/CA}
\draw[draw=black,fill=color2, postaction={pattern=dots},draw opacity=1] (axis cs:0.6,0) rectangle (axis cs:0.866666666666667,0.0148666666666667);
\draw[fill=color2, postaction={pattern=dots},draw opacity=1] (axis cs:1.6,0) rectangle (axis cs:1.866666666666667,0.0313);
\draw[fill=color2, postaction={pattern=dots},draw opacity=1] (axis cs:2.6,0) rectangle (axis cs:2.866666666666667,0.671633333333333);
\draw[fill=color2, postaction={pattern=dots},draw opacity=1] (axis cs:3.6,0) rectangle (axis cs:3.866666666666667,0.7325);
\draw[fill=color2, postaction={pattern=dots},draw opacity=1] (axis cs:4.6,0) rectangle (axis cs:4.866666666666667,0.7924);
\draw[fill=color2, postaction={pattern=dots},draw opacity=1] (axis cs:5.6,0) rectangle (axis cs:5.866666666666667,0.8587);
\draw[fill=color2, postaction={pattern=dots},draw opacity=1] (axis cs:6.6,0) rectangle (axis cs:6.866666666666667,0.918366666666667);
\draw[fill=color2, postaction={pattern=dots},draw opacity=1] (axis cs:7.6,0) rectangle (axis cs:7.866666666666667,0.958266666666667);

\addplot [line width=1.08pt, white!26.0!black, forget plot]
table {%
0.266666666666667 0.23363
0.266666666666667 0.46794
};
\addplot [line width=1.08pt, white!26.0!black, forget plot]
table {%
1.26666666666667 0.5276175
1.26666666666667 0.708566666666666
};
\addplot [line width=1.08pt, white!26.0!black, forget plot]
table {%
2.26666666666667 0.967230833333333
2.26666666666667 0.982001666666667
};
\addplot [line width=1.08pt, white!26.0!black, forget plot]
table {%
3.26666666666667 0.998833333333333
3.26666666666667 0.9998
};
\addplot [line width=1.08pt, white!26.0!black, forget plot]
table {%
4.26666666666667 0.999533333333333
4.26666666666667 0.999933333333333
};
\addplot [line width=1.08pt, white!26.0!black, forget plot]
table {%
5.26666666666667 0.999433333333333
5.26666666666667 0.9999
};
\addplot [line width=1.08pt, white!26.0!black, forget plot]
table {%
6.26666666666667 0.992333333333333
6.26666666666667 0.999833333333333
};
\addplot [line width=1.08pt, white!26.0!black, forget plot]
table {%
7.26666666666667 0.999433333333334
7.26666666666667 0.999933333333333
};
\addplot [line width=1.08pt, white!26.0!black, forget plot]
table {%
8.26666666666667 0.9999
8.26666666666667 1
};
\addplot [line width=1.08pt, white!26.0!black, forget plot]
table {%
0 0.00483333333333333
0 0.00676666666666667
};
\addplot [line width=1.08pt, white!26.0!black, forget plot]
table {%
1 0.0093
1 0.0114666666666667
};
\addplot [line width=1.08pt, white!26.0!black, forget plot]
table {%
2 0.0214666666666667
2 0.0246
};
\addplot [line width=1.08pt, white!26.0!black, forget plot]
table {%
3 0.6571
3 0.6708
};
\addplot [line width=1.08pt, white!26.0!black, forget plot]
table {%
4 0.720530833333333
4 0.7304675
};
\addplot [line width=1.08pt, white!26.0!black, forget plot]
table {%
5 0.779599166666667
5 0.791934166666667
};
\addplot [line width=1.08pt, white!26.0!black, forget plot]
table {%
6 0.8433325
6 0.853234166666667
};
\addplot [line width=1.08pt, white!26.0!black, forget plot]
table {%
7 0.913933333333334
7 0.919135
};
\addplot [line width=1.08pt, white!26.0!black, forget plot]
table {%
8 0.9556325
8 0.9610025
};
\addplot [line width=1.08pt, white!26.0!black, forget plot]
table {%
-0.266666666666667 0.00546666666666667
-0.266666666666667 0.00713333333333333
};
\addplot [line width=1.08pt, white!26.0!black, forget plot]
table {%
0.733333333333333 0.0133991666666667
0.733333333333333 0.0164
};
\addplot [line width=1.08pt, white!26.0!black, forget plot]
table {%
1.733333333333333 0.0286991666666667
1.733333333333333 0.0340675
};
\addplot [line width=1.08pt, white!26.0!black, forget plot]
table {%
2.733333333333333 0.665565
2.733333333333333 0.6779675
};
\addplot [line width=1.08pt, white!26.0!black, forget plot]
table {%
3.733333333333333 0.727465833333333
3.733333333333333 0.738035
};
\addplot [line width=1.08pt, white!26.0!black, forget plot]
table {%
4.733333333333333 0.787666666666667
4.733333333333333 0.796366666666666
};
\addplot [line width=1.08pt, white!26.0!black, forget plot]
table {%
5.733333333333333 0.855299166666667
5.733333333333333 0.862535
};
\addplot [line width=1.08pt, white!26.0!black, forget plot]
table {%
6.733333333333333 0.915066666666667
6.733333333333333 0.921200833333333
};
\addplot [line width=1.08pt, white!26.0!black, forget plot]
table {%
7.733333333333333 0.955365833333333
7.733333333333333 0.960935
};
\end{axis}

\end{tikzpicture}

%% file: plots/queue_hidden.tex
\begin{tikzpicture}

\begin{axis}[
width=\FIGWIDTH,
height=5cm,
legend columns=3, 
legend style={fill opacity=1.8, draw opacity=1, text opacity=1, anchor=south, draw=black, at={(0.5, 1.1)}},
tick align=outside,
tick pos=both,
x grid style={white!69.01960784313725!black},
xlabel={packet generation rate $\delta$ [packets/s]},
xmin=-0.5, xmax=8.5,
xtick style={draw=black,color=black},
y grid style={white!69.01960784313725!black},
ylabel={Avg. \# packets in queue},
ymin=0, ymax=8.36356030688107,
ytick style={draw=black,color=black},
xtick={0,1,2,3,4,5,6,7,8,9},
xticklabels={100, 50, 25, 10, 8, 6, 4, 2, 1},
x dir=reverse
]

\draw[fill=color0,draw opacity=1] (axis cs:0.13333333333333,0) rectangle (axis cs:0.4,6.32248632220512);
\addlegendimage{ybar,ybar legend,fill=color0,draw opacity=1};
\addlegendentry{QMA}
\draw[fill=color0,draw opacity=1] (axis cs:1.13333333333333,0) rectangle (axis cs:1.4,5.71117776459626);
\draw[fill=color0,draw opacity=1] (axis cs:2.13333333333333,0) rectangle (axis cs:2.4,2.73774667758643);
\draw[fill=color0,draw opacity=1] (axis cs:3.13333333333333,0) rectangle (axis cs:3.4,1.57148245491862);
\draw[fill=color0,draw opacity=1] (axis cs:4.13333333333333,0) rectangle (axis cs:4.4,1.48234590942832);
\draw[fill=color0,draw opacity=1] (axis cs:5.13333333333333,0) rectangle (axis cs:5.4,1.41054014615385);
\draw[fill=color0,draw opacity=1] (axis cs:6.13333333333333,0) rectangle (axis cs:6.4,1.33889735238794);
\draw[fill=color0,draw opacity=1] (axis cs:7.13333333333333,0) rectangle (axis cs:7.4,1.23540746866168);
\draw[fill=color0,draw opacity=1] (axis cs:8.13333333333333,0) rectangle (axis cs:8.4,1.18061283306533);
\draw[fill=color1, postaction={pattern=north east lines},draw opacity=1] (axis cs:-0.133333333333333,0) rectangle (axis cs:0.133333333333333,7.96479299426636);
\addlegendimage{ybar,ybar legend,fill=color1, postaction={pattern=north east lines},draw opacity=1};
\addlegendentry{slotted CSMA/CA}

\draw[fill=color1, postaction={pattern=north east lines},draw opacity=1] (axis cs:0.866666666666667,0) rectangle (axis cs:1.13333333333333,7.85540284026921);
\draw[fill=color1, postaction={pattern=north east lines},draw opacity=1] (axis cs:1.86666666666667,0) rectangle (axis cs:2.13333333333333,7.23657152650764);
\draw[fill=color1, postaction={pattern=north east lines},draw opacity=1] (axis cs:2.86666666666667,0) rectangle (axis cs:3.13333333333333,1.48536686706583);
\draw[fill=color1, postaction={pattern=north east lines},draw opacity=1] (axis cs:3.86666666666667,0) rectangle (axis cs:4.13333333333333,1.32961706616811);
\draw[fill=color1, postaction={pattern=north east lines},draw opacity=1] (axis cs:4.86666666666667,0) rectangle (axis cs:5.13333333333333,1.21793745420722);
\draw[fill=color1, postaction={pattern=north east lines},draw opacity=1] (axis cs:5.86666666666667,0) rectangle (axis cs:6.13333333333333,1.12985760344718);
\draw[fill=color1, postaction={pattern=north east lines},draw opacity=1] (axis cs:6.86666666666667,0) rectangle (axis cs:7.13333333333333,1.06014912855322);
\draw[fill=color1, postaction={pattern=north east lines},draw opacity=1] (axis cs:7.86666666666667,0) rectangle (axis cs:8.13333333333333,1.02840976043833);

\draw[fill=color2, postaction={pattern=dots},draw opacity=1] (axis cs:-0.4,0) rectangle (axis cs:-0.13333333333333333,7.94695970935154);
\addlegendimage{ybar,ybar legend,fill=color2, postaction={pattern=dots},draw opacity=1};
\addlegendentry{unslotted CSMA/CA}
\draw[fill=color2, postaction={pattern=dots},draw opacity=1] (axis cs:0.6,0) rectangle (axis cs:0.866666666666667,7.79617223696858);
\draw[fill=color2, postaction={pattern=dots},draw opacity=1] (axis cs:1.6,0) rectangle (axis cs:1.866666666666667,6.82405431960464);
\draw[fill=color2, postaction={pattern=dots},draw opacity=1] (axis cs:2.6,0) rectangle (axis cs:2.866666666666667,1.42283425471658);
\draw[fill=color2, postaction={pattern=dots},draw opacity=1] (axis cs:3.6,0) rectangle (axis cs:3.866666666666667,1.29623746754527);
\draw[fill=color2, postaction={pattern=dots},draw opacity=1] (axis cs:4.6,0) rectangle (axis cs:4.866666666666667,1.20181136094926);
\draw[fill=color2, postaction={pattern=dots},draw opacity=1] (axis cs:5.6,0) rectangle (axis cs:5.866666666666667,1.12193641474896);
\draw[fill=color2, postaction={pattern=dots},draw opacity=1] (axis cs:6.6,0) rectangle (axis cs:6.866666666666667,1.0565884104117);
\draw[fill=color2, postaction={pattern=dots},draw opacity=1] (axis cs:7.6,0) rectangle (axis cs:7.866666666666667,1.0279465044219);
\addplot [line width=1.08pt, white!26.0!black, forget plot]
table {%
0.266666666666667 5.79502457314056
0.266666666666667 6.83362889955487
};
\addplot [line width=1.08pt, white!26.0!black, forget plot]
table {%
1.266666666666667 5.49036544320083
1.266666666666667 5.94398843964874
};
\addplot [line width=1.08pt, white!26.0!black, forget plot]
table {%
2.266666666666667 2.68031051930486
2.266666666666667 2.79396503193204
};
\addplot [line width=1.08pt, white!26.0!black, forget plot]
table {%
3.266666666666667 1.54607781300726
3.266666666666667 1.59910532953102
};
\addplot [line width=1.08pt, white!26.0!black, forget plot]
table {%
4.266666666666667 1.44918671782079
4.266666666666667 1.52160197963042
};
\addplot [line width=1.08pt, white!26.0!black, forget plot]
table {%
5.266666666666667 1.38353539450816
5.266666666666667 1.43699120211209
};
\addplot [line width=1.08pt, white!26.0!black, forget plot]
table {%
6.266666666666667 1.30923176609598
6.266666666666667 1.36734077341929
};
\addplot [line width=1.08pt, white!26.0!black, forget plot]
table {%
7.266666666666667 1.21784074451607
7.266666666666667 1.2557427432715
};
\addplot [line width=1.08pt, white!26.0!black, forget plot]
table {%
8.266666666666667 1.16380534798088
8.266666666666667 1.1972527856041
};
\addplot [line width=1.08pt, white!26.0!black, forget plot]
table {%
0 7.96424556605036
0 7.96529553036293
};
\addplot [line width=1.08pt, white!26.0!black, forget plot]
table {%
1 7.85432277564136
1 7.85654683592548
};
\addplot [line width=1.08pt, white!26.0!black, forget plot]
table {%
2 7.22826972334539
2 7.244233448507
};
\addplot [line width=1.08pt, white!26.0!black, forget plot]
table {%
3 1.48018771331729
3 1.49035395113736
};
\addplot [line width=1.08pt, white!26.0!black, forget plot]
table {%
4 1.32645202131003
4 1.33270015405092
};
\addplot [line width=1.08pt, white!26.0!black, forget plot]
table {%
5 1.21584017009709
5 1.22015624242741
};
\addplot [line width=1.08pt, white!26.0!black, forget plot]
table {%
6 1.12784965945709
6 1.13193261034451
};
\addplot [line width=1.08pt, white!26.0!black, forget plot]
table {%
7 1.05873025797491
7 1.06163977983674
};
\addplot [line width=1.08pt, white!26.0!black, forget plot]
table {%
8 1.02700302802473
8 1.0297637462564
};
\addplot [line width=1.08pt, white!26.0!black, forget plot]
table {%
-0.266666666666667 7.94647068528251
-0.266666666666667 7.94743225555012
};
\addplot [line width=1.08pt, white!26.0!black, forget plot]
table {%
0.733333333333333 7.79503217542357
0.733333333333333 7.79728419695451
};
\addplot [line width=1.08pt, white!26.0!black, forget plot]
table {%
1.733333333333333 6.80979404361508
1.733333333333333 6.8371623234149
};
\addplot [line width=1.08pt, white!26.0!black, forget plot]
table {%
2.733333333333333 1.41905493166775
2.733333333333333 1.42679775927366
};
\addplot [line width=1.08pt, white!26.0!black, forget plot]
table {%
3.733333333333333 1.29312187310312
3.733333333333333 1.29939711130582
};
\addplot [line width=1.08pt, white!26.0!black, forget plot]
table {%
4.733333333333333 1.19967219895807
4.733333333333333 1.20383340985413
};
\addplot [line width=1.08pt, white!26.0!black, forget plot]
table {%
5.733333333333333 1.12014376802551
5.733333333333333 1.12391594580112
};
\addplot [line width=1.08pt, white!26.0!black, forget plot]
table {%
6.733333333333333 1.05538120746108
6.733333333333333 1.05782346641253
};
\addplot [line width=1.08pt, white!26.0!black, forget plot]
table {%
7.733333333333333 1.02676221695918
7.733333333333333 1.02919639573114
};
\end{axis}

\end{tikzpicture}

%% file: plots/delay_hidden.tex
\begin{tikzpicture}

\begin{axis}[
width=\FIGWIDTH,
height=\FIGHEIGHT,
legend columns=3,
legend style={fill opacity=1.8, draw opacity=1, text opacity=1, anchor=south, draw=black, at={(0.5, 1.1)}},
tick align=outside,
tick pos=both,
x grid style={white!69.01960784313725!black},
xlabel={packet generation rate $\delta$ [packets/s]},
xmin=-0.6, xmax=8.4,
xtick style={color=black},
y grid style={white!69.01960784313725!black},
ylabel={avg. end-to-end delay [s]},
ymin=0, ymax=0.415406681707867,
ytick style={color=black},
xtick={-0.1,0.9,1.9,2.9,3.9,4.9,5.9,6.9,7.9,8.9},
xticklabels={100, 50, 25, 10, 8, 6, 4, 2, 1},
x dir=reverse
]
\draw[fill=color0,draw opacity=1] (axis cs:0,0) rectangle (axis cs:0.2,0.296088207862746);
\addlegendimage{ybar,ybar legend,fill=color0,draw opacity=1};
\addlegendentry{QMA}

\draw[fill=color0,draw opacity=1] (axis cs:1,0) rectangle (axis cs:1.2,0.299167379724972);
\draw[fill=color0,draw opacity=1] (axis cs:2,0) rectangle (axis cs:2.2,0.23390847855277);
\draw[fill=color0,draw opacity=1] (axis cs:3,0) rectangle (axis cs:3.2,0.0217306272514563);
\draw[fill=color0,draw opacity=1] (axis cs:4,0) rectangle (axis cs:4.2,0.0243796541250293);
\draw[fill=color0,draw opacity=1] (axis cs:5,0) rectangle (axis cs:5.2,0.0255331866918538);
\draw[fill=color0,draw opacity=1] (axis cs:6,0) rectangle (axis cs:6.2,0.0278227056626865);
\draw[fill=color0,draw opacity=1] (axis cs:7,0) rectangle (axis cs:7.2,0.0296238755215009);
\draw[fill=color0,draw opacity=1] (axis cs:8,0) rectangle (axis cs:8.2,0.0313417695411842);

\draw[fill=color1, postaction={pattern=north east lines},draw opacity=1] (axis cs:-0.2,0) rectangle (axis cs:0.0,0.393165154062516);
\addlegendimage{ybar,ybar legend,fill=color1, postaction={pattern=north east lines},draw opacity=1};
\addlegendentry{slotted CSMA/CA}

\draw[fill=color1, postaction={pattern=north east lines},draw opacity=1] (axis cs:0.8,0) rectangle (axis cs:1,0.385305635156364);
\draw[fill=color1, postaction={pattern=north east lines},draw opacity=1] (axis cs:1.8,0) rectangle (axis cs:2,0.367444939752644);
\draw[fill=color1, postaction={pattern=north east lines},draw opacity=1] (axis cs:2.8,0) rectangle (axis cs:3,0.0193165154062516);
\draw[fill=color1, postaction={pattern=north east lines},draw opacity=1] (axis cs:3.8,0) rectangle (axis cs:4,0.0184459601520701);
\draw[fill=color1, postaction={pattern=north east lines},draw opacity=1] (axis cs:4.8,0) rectangle (axis cs:5,0.0208863079675972);
\draw[fill=color1, postaction={pattern=north east lines},draw opacity=1] (axis cs:5.8,0) rectangle (axis cs:6,0.0229267505988198);
\draw[fill=color1, postaction={pattern=north east lines},draw opacity=1] (axis cs:6.8,0) rectangle (axis cs:7,0.0234454071145626);
\draw[fill=color1, postaction={pattern=north east lines},draw opacity=1] (axis cs:7.8,0) rectangle (axis cs:8,0.022209673500483);
\draw[fill=color2, postaction={pattern=dots},draw opacity=1] (axis cs:-0.4,0) rectangle (axis cs:-0.2,0.390043391769407);
\addlegendimage{ybar,ybar legend,fill=color2, postaction={pattern=dots},draw opacity=1};
\addlegendentry{unslotted CSMA/CA}

\draw[fill=color2, postaction={pattern=dots},draw opacity=1] (axis cs:0.6,0) rectangle (axis cs:0.8,0.384536450234131);
\draw[fill=color2, postaction={pattern=dots},draw opacity=1] (axis cs:1.6,0) rectangle (axis cs:1.8,0.364963732930881);
\draw[fill=color2, postaction={pattern=dots},draw opacity=1] (axis cs:2.6,0) rectangle (axis cs:2.8,0.0171064490889092);
\draw[fill=color2, postaction={pattern=dots},draw opacity=1] (axis cs:3.6,0) rectangle (axis cs:3.8,0.0186847098404815);
\draw[fill=color2, postaction={pattern=dots},draw opacity=1] (axis cs:4.6,0) rectangle (axis cs:4.8,0.0199108817532535);
\draw[fill=color2, postaction={pattern=dots},draw opacity=1] (axis cs:5.6,0) rectangle (axis cs:5.8,0.0239571080160205);
\draw[fill=color2, postaction={pattern=dots},draw opacity=1] (axis cs:6.6,0) rectangle (axis cs:6.8,0.0237924912167524);
\draw[fill=color2, postaction={pattern=dots},draw opacity=1] (axis cs:7.6,0) rectangle (axis cs:7.8,0.0242545394806147);

\addplot [line width=1.08pt, white!26.0!black, forget plot]
table {%
0.1 0.218073742370197
0.1 0.387744674095517
};
\addplot [line width=1.08pt, white!26.0!black, forget plot]
table {%
1.1 0.245721704966891
1.1 0.35281544657183
};
\addplot [line width=1.08pt, white!26.0!black, forget plot]
table {%
2.1 0.186747059098607
2.1 0.278445326720308
};
\addplot [line width=1.08pt, white!26.0!black, forget plot]
table {%
3.1 0.0188663568276289
3.1 0.0250745888278639
};
\addplot [line width=1.08pt, white!26.0!black, forget plot]
table {%
4.1 0.0213427308641419
4.1 0.0272456335836292
};
\addplot [line width=1.08pt, white!26.0!black, forget plot]
table {%
5.1 0.0229956100735517
5.1 0.0278837687903967
};
\addplot [line width=1.08pt, white!26.0!black, forget plot]
table {%
6.1 0.0259498509487461
6.1 0.0294206894997158
};
\addplot [line width=1.08pt, white!26.0!black, forget plot]
table {%
7.1 0.0272831895601694
7.1 0.032015203978705
};
\addplot [line width=1.08pt, white!26.0!black, forget plot]
table {%
8.1 0.0297515235165189
8.1 0.0331458634148178
};
\addplot [line width=1.08pt, white!26.0!black, forget plot]
table {%
-0.1 0.390933624629823
-0.1 0.395337819253151
};
\addplot [line width=1.08pt, white!26.0!black, forget plot]
table {%
0.9 0.383391673824531
0.9 0.386989543958369
};
\addplot [line width=1.08pt, white!26.0!black, forget plot]
table {%
1.9 0.365269958000936
1.9 0.369606085748497
};
\addplot [line width=1.08pt, white!26.0!black, forget plot]
table {%
2.9 0.0190738799195121
2.9 0.019562541115035
};
\addplot [line width=1.08pt, white!26.0!black, forget plot]
table {%
3.9 0.0163213197706288
3.9 0.0204474645831563
};
\addplot [line width=1.08pt, white!26.0!black, forget plot]
table {%
4.9 0.0188952386333744
4.9 0.0227408054705752
};
\addplot [line width=1.08pt, white!26.0!black, forget plot]
table {%
5.9 0.0204707223345652
5.9 0.0248393239227313
};
\addplot [line width=1.08pt, white!26.0!black, forget plot]
table {%
6.9 0.0214759282031918
6.9 0.0248685368048561
};
\addplot [line width=1.08pt, white!26.0!black, forget plot]
table {%
7.9 0.0198070289473938
7.9 0.024488852459111
};
\addplot [line width=1.08pt, white!26.0!black, forget plot]
table {%
-0.3 0.387298287113904
-0.3 0.39296205149464
};
\addplot [line width=1.08pt, white!26.0!black, forget plot]
table {%
0.7 0.382682702406257
0.7 0.386161173067252
};
\addplot [line width=1.08pt, white!26.0!black, forget plot]
table {%
1.7 0.363058183350935
1.7 0.366925946157649
};
\addplot [line width=1.08pt, white!26.0!black, forget plot]
table {%
2.7 0.0156284298187748
2.7 0.0185959724609759
};
\addplot [line width=1.08pt, white!26.0!black, forget plot]
table {%
3.7 0.0167582217529906
3.7 0.0206656465201436
};
\addplot [line width=1.08pt, white!26.0!black, forget plot]
table {%
4.7 0.0173529137173722
4.7 0.0224162731380497
};
\addplot [line width=1.08pt, white!26.0!black, forget plot]
table {%
5.7 0.0220029688219431
5.7 0.0252645065975625
};
\addplot [line width=1.08pt, white!26.0!black, forget plot]
table {%
6.7 0.0220532968685346
6.7 0.0252583689016094
};
\addplot [line width=1.08pt, white!26.0!black, forget plot]
table {%
7.7 0.0230526308928042
7.7 0.0252303768164377
};
\end{axis}

\end{tikzpicture}

%% file: plots/slot_utilization_hidden_1a.tex
\begin{tikzpicture}

\begin{axis}[
	legend style={at={(0.5,1.1)},anchor=south},
	legend columns=2,
    legend entries={Node C TX, Node A CCA},
    height=4cm,
    width=\textwidth,
    ymin = 0,
    ymax = 3,
    xmin = 0,
    xmax = 53,
    xlabel={Subslot id},
    ylabel={Node},
    ytick={1,2},
    yticklabels={A,C},
    minor tick num=1,
    grid=both
]
\addplot[only marks, mark=square*, mark options={fill=color1}]
table[] {
    x y 
    18 2 
};
\addplot[only marks, mark=*, mark options={fill=color0}]
table[] {
	x y 
	36 1  
};
\end{axis}

\end{tikzpicture}

%% file: plots/slot_utilization_hidden_1b.tex
\begin{tikzpicture}

\begin{axis}[
	legend style={at={(0.5,1.1)},anchor=south},
	legend columns=3,
    legend entries={Node C TX, Node A CCA, Node A TX},
    height=4cm,
    width=\textwidth,
    ymin = 0,
    ymax = 3,
    xmin = 0,
    xmax = 53,
    xlabel={Subslot id},
    ylabel={Node},
    ytick={1,2},
    yticklabels={A,C},
    minor tick num=1,
    grid=both
]
\addplot[only marks, mark=square*, mark options={fill=color1}]
table[] {
	x y 
	18 2 
};
\addplot[only marks, mark=*, mark options={fill=color0}]
table[] {
	x y 
	36 1  
};
\addplot[only marks, mark=square*, mark options={fill=color0}]
table[] {
	x y 
	14 1  
};
\end{axis}

\end{tikzpicture}

%% file: plots/slot_utilization_hidden_10a.tex
\begin{tikzpicture}

\begin{axis}[
	legend style={at={(0.5,1.1)},anchor=south},
    legend entries={Node C TX},
    height=4cm,
    width=\textwidth,
    ymin = 0,
    ymax = 3,
    xmin = 0,
    xmax = 53,
    xlabel={Subslot id},
    ylabel={Node},
    ytick={1,2},
    yticklabels={A,C},
    minor tick num=1,
    grid=both
]
\addplot[only marks, mark=square*, mark options={fill=color1}]
table[] {
    x y 
    7 2 
    22 2 
    33 2  
};
\end{axis}

\end{tikzpicture}

%% file: plots/slot_utilization_hidden_10b.tex
\begin{tikzpicture}

\begin{axis}[
	legend columns=2,
	legend style={at={(0.5,1.1)},anchor=south},
    legend entries={Node C TX, Node A TX},
    height=4cm,
    width=\textwidth,
    ymin = 0,
    ymax = 3,
    xmin = 0,
    xmax = 53,
    xlabel={Subslot id},
    ylabel={Node},
    ytick={1,2},
    yticklabels={A,C},
    minor tick num=1,
    grid=both
]
\addplot[only marks, mark=square*, mark options={fill=color1}]
table[] {
    x y 
    7 2 
    28 2 
    33 2 
    44 2  
};
\addplot[only marks, mark=square*, mark options={fill=color0}]
table[] {
	x y 
	17 1
	24 1 
	27 1
	47 1
	48 1   
};  
\end{axis}

\end{tikzpicture}

%% file: plots/slot_utilization_hidden_100a.tex
\begin{tikzpicture}

\begin{axis}[
	legend style={at={(0.5,1.1)},anchor=south},
	legend columns=2,
    legend entries={Node C CCA, Node C TX},
    height=4cm,
    width=\textwidth,
    ymin = 0,
    ymax = 3,
    xmin = 0,
    xmax = 53,
    xlabel={Subslot id},
    ylabel={Node},
    ytick={1,2},
    yticklabels={A,C},
    minor tick num=1,
    grid=both
]

\addplot[only marks, mark=*, mark options={fill=color1}]
table[] {
	x y 
	11 2 
	12 2 
	13 2
	34 2   
};
\addplot[only marks, mark=square*, mark options={fill=color1}]
table[] {
    x y 
    8 2 
    14 2 
    15 2
    17 2 
    20 2 
    22 2 
    26 2 
    29 2 
    30 2 
    33 2 
    41 2 
    42 2 
    48 2   
};
\end{axis}

\end{tikzpicture}

%% file: plots/slot_utilization_hidden_100b.tex
\begin{tikzpicture}

\begin{axis}[
	legend style={at={(0.5,1.1)},anchor=south},
	legend columns=2,
    legend entries={Node C CCA, Node C TX, Node A CCA, Node A TX},
    height=4cm,
    width=\textwidth,
    ymin = 0,
    ymax = 3,
    xmin = 0,
    xmax = 53,
    xlabel={Subslot id},
    ylabel={Node},
    ytick={1,2},
    yticklabels={A,C},
    minor tick num=1,
    grid=both
]

\addplot[only marks, mark=*, mark options={fill=color1}]
table[] {
	x y 
	13 2 
	23 2 
	24 2 
};
\addplot[only marks, mark=square*, mark options={fill=color1}]
table[] {
    x y 
    6 2 
    7 2 
    8 2 
    10 2 
    11 2 
    12 2 
    14 2 
    15 2 
    16 2 
    17 2 
    18 2 
    19 2 
    20 2 
    21 2 
    22 2 
    25 2 
    26 2   
};

\addplot[only marks, mark=*, mark options={fill=color0}]
table[] {
	x y 
	32 1 
	35 1 
	48 1 
};
\addplot[only marks, mark=square*, mark options={fill=color0}]
table[] {
	x y 
	30 1 
	36 1 
	37 1 
	39 1 
	40 1 
	41 1 
	42 1 
	43 1 
	44 1 
	45 1 
	46 1 
	47 1 
	49 1    
};
\end{axis}

\end{tikzpicture}

%% file: plots/iotlab_tree_prr.tex
\begin{tikzpicture}

\begin{axis}[
width=.65*\FIGWIDTH,
height=5cm,
legend cell align={left},
legend columns=2,
legend style={fill opacity=1.8, draw opacity=1, text opacity=1, draw=white!80.0!black, anchor=south, at={(0.5,1.1)}},
tick align=outside,
tick pos=both,
x grid style={white!69.01960784313725!black},
xlabel={node id},
xmin=-0.5, xmax=8.5,
xtick style={draw=black,color=black},
xtick={0,1,2,3,4,5,6,7,8},
xticklabels={15,18,19,2,36,41,59,63,64},
y grid style={white!69.01960784313725!black},
ylabel={packet delivery ratio},
ymin=0, ymax=1.049825,
ytick style={draw=black,color=black}
]
\draw[draw=black,fill=color0,draw opacity=1] (axis cs:-0.4,0) rectangle (axis cs:0,0.9685);
\addlegendimage{ybar,ybar legend,draw=black,fill=color0,draw opacity=1};
\addlegendentry{QMA}

\draw[draw=black,fill=color0,draw opacity=1] (axis cs:0.6,0) rectangle (axis cs:1,0.998);
\draw[draw=black,fill=color0,draw opacity=1] (axis cs:1.6,0) rectangle (axis cs:2,0.867);
\draw[draw=black,fill=color0,draw opacity=1] (axis cs:2.6,0) rectangle (axis cs:3,0.902833333333333);
\draw[draw=black,fill=color0,draw opacity=1] (axis cs:3.6,0) rectangle (axis cs:4,0.7265);
\draw[draw=black,fill=color0,draw opacity=1] (axis cs:4.6,0) rectangle (axis cs:5,0.9415);
\draw[draw=black,fill=color0,draw opacity=1] (axis cs:5.6,0) rectangle (axis cs:6,0.950666666666667);
\draw[draw=black,fill=color0,draw opacity=1] (axis cs:6.6,0) rectangle (axis cs:7,0.8646);
\draw[draw=black,fill=color0,draw opacity=1] (axis cs:7.6,0) rectangle (axis cs:8,0.828);
\draw[draw=black,fill=color1, postaction={pattern=north east lines},draw opacity=1] (axis cs:0,0) rectangle (axis cs:0.4,0.5984);
\addlegendimage{ybar,ybar legend,draw=black,fill=color1, postaction={pattern=north east lines},draw opacity=1};
\addlegendentry{unslotted CSMA/CA}

\draw[draw=black,fill=color1, postaction={pattern=north east lines},draw opacity=1] (axis cs:1,0) rectangle (axis cs:1.4,0.7032);
\draw[draw=black,fill=color1, postaction={pattern=north east lines},draw opacity=1] (axis cs:2,0) rectangle (axis cs:2.4,0.5382);
\draw[draw=black,fill=color1, postaction={pattern=north east lines},draw opacity=1] (axis cs:3,0) rectangle (axis cs:3.4,0.5124);
\draw[draw=black,fill=color1, postaction={pattern=north east lines},draw opacity=1] (axis cs:4,0) rectangle (axis cs:4.4,0.5428);
\draw[draw=black,fill=color1, postaction={pattern=north east lines},draw opacity=1] (axis cs:5,0) rectangle (axis cs:5.4,0.5944);
\draw[draw=black,fill=color1, postaction={pattern=north east lines},draw opacity=1] (axis cs:6,0) rectangle (axis cs:6.4,0.4766);
\draw[draw=black,fill=color1, postaction={pattern=north east lines},draw opacity=1] (axis cs:7,0) rectangle (axis cs:7.4,0.7104);
\draw[draw=black,fill=color1, postaction={pattern=north east lines},draw opacity=1] (axis cs:8,0) rectangle (axis cs:8.4,0.541);
\addplot [line width=1.08pt, white!26.0!black, forget plot]
table {%
-0.2 0.917
-0.2 0.999833333333333
};
\addplot [line width=1.08pt, white!26.0!black, forget plot]
table {%
0.8 0.9968
0.8 0.9992
};
\addplot [line width=1.08pt, white!26.0!black, forget plot]
table {%
1.8 0.604666666666667
1.8 0.999333333333333
};
\addplot [line width=1.08pt, white!26.0!black, forget plot]
table {%
2.8 0.713495833333333
2.8 0.999
};
\addplot [line width=1.08pt, white!26.0!black, forget plot]
table {%
3.8 0.396833333333333
3.8 0.989
};
\addplot [line width=1.08pt, white!26.0!black, forget plot]
table {%
4.8 0.844833333333333
4.8 0.997166666666667
};
\addplot [line width=1.08pt, white!26.0!black, forget plot]
table {%
5.8 0.889325
5.8 0.9968375
};
\addplot [line width=1.08pt, white!26.0!black, forget plot]
table {%
6.8 0.6156
6.8 0.996005
};
\addplot [line width=1.08pt, white!26.0!black, forget plot]
table {%
7.8 0.4955
7.8 0.997166666666667
};
\addplot [line width=1.08pt, white!26.0!black, forget plot]
table {%
0.2 0.55173
0.2 0.6478
};
\addplot [line width=1.08pt, white!26.0!black, forget plot]
table {%
1.2 0.6516
1.2 0.7548
};
\addplot [line width=1.08pt, white!26.0!black, forget plot]
table {%
2.2 0.4312
2.2 0.6172
};
\addplot [line width=1.08pt, white!26.0!black, forget plot]
table {%
3.2 0.467
3.2 0.5618
};
\addplot [line width=1.08pt, white!26.0!black, forget plot]
table {%
4.2 0.517
4.2 0.569645
};
\addplot [line width=1.08pt, white!26.0!black, forget plot]
table {%
5.2 0.4862
5.2 0.66
};
\addplot [line width=1.08pt, white!26.0!black, forget plot]
table {%
6.2 0.412
6.2 0.5478
};
\addplot [line width=1.08pt, white!26.0!black, forget plot]
table {%
7.2 0.7004
7.2 0.7204
};
\addplot [line width=1.08pt, white!26.0!black, forget plot]
table {%
8.2 0.5024
8.2 0.5796
};
\end{axis}

\end{tikzpicture}

%% file: plots/iotlab_star_prr.tex
\begin{tikzpicture}

\begin{axis}[
height=5cm,
width=\FIGWIDTH,
legend columns=3,
legend cell align={left},
legend style={fill opacity=1.8, draw opacity=1, text opacity=1, at={(0.5,1.1)}, anchor=south, draw=white!80.0!black},
tick align=outside,
tick pos=both,
x grid style={white!69.01960784313725!black},
xlabel={node id},
xmin=-0.5, xmax=15.5,
xtick style={draw=black,color=black},
xtick={0,1,2,3,4,5,6,7,8,9,10,11,12,13,14,15},
xticklabels={10,2,20,24,30,38,4,48,52,54,56,58,6,60,62,8},
y grid style={white!69.01960784313725!black},
ylabel={packet delivery ratio},
ymin=0, ymax=0.832810125,
ytick style={draw=black,color=black}
]
\draw[draw=black,fill=color0,draw opacity=1] (axis cs:-0.4,0) rectangle (axis cs:0,0.4912);
\addlegendimage{ybar,ybar legend,draw=black,fill=color0,draw opacity=1};
\addlegendentry{QMA}

\draw[draw=black,fill=color0,draw opacity=1] (axis cs:0.6,0) rectangle (axis cs:1,0.6576);
\draw[draw=black,fill=color0,draw opacity=1] (axis cs:1.6,0) rectangle (axis cs:2,0.496888888888889);
\draw[draw=black,fill=color0,draw opacity=1] (axis cs:2.6,0) rectangle (axis cs:3,0.580777777777778);
\draw[draw=black,fill=color0,draw opacity=1] (axis cs:3.6,0) rectangle (axis cs:4,0.5796);
\draw[draw=black,fill=color0,draw opacity=1] (axis cs:4.6,0) rectangle (axis cs:5,0.6234);
\draw[draw=black,fill=color0,draw opacity=1] (axis cs:5.6,0) rectangle (axis cs:6,0.5147);
\draw[draw=black,fill=color0,draw opacity=1] (axis cs:6.6,0) rectangle (axis cs:7,0.52575);
\draw[draw=black,fill=color0,draw opacity=1] (axis cs:7.6,0) rectangle (axis cs:8,0.6017);
\draw[draw=black,fill=color0,draw opacity=1] (axis cs:8.6,0) rectangle (axis cs:9,0.52875);
\draw[draw=black,fill=color0,draw opacity=1] (axis cs:9.6,0) rectangle (axis cs:10,0.6144);
\draw[draw=black,fill=color0,draw opacity=1] (axis cs:10.6,0) rectangle (axis cs:11,0.7038);
\draw[draw=black,fill=color0,draw opacity=1] (axis cs:11.6,0) rectangle (axis cs:12,0.5179);
\draw[draw=black,fill=color0,draw opacity=1] (axis cs:12.6,0) rectangle (axis cs:13,0.6852);
\draw[draw=black,fill=color0,draw opacity=1] (axis cs:13.6,0) rectangle (axis cs:14,0.5425);
\draw[draw=black,fill=color0,draw opacity=1] (axis cs:14.6,0) rectangle (axis cs:15,0.648555555555556);
\draw[draw=black,fill=color1, postaction={pattern=north east lines},draw opacity=1] (axis cs:0,0) rectangle (axis cs:0.4,0.4357);
\addlegendimage{ybar,ybar legend,draw=black,fill=color1, postaction={pattern=north east lines},draw opacity=1};
\addlegendentry{unslotted CSMA/CA}

\draw[draw=black,fill=color1, postaction={pattern=north east lines},draw opacity=1] (axis cs:1,0) rectangle (axis cs:1.4,0.4602);
\draw[draw=black,fill=color1, postaction={pattern=north east lines},draw opacity=1] (axis cs:2,0) rectangle (axis cs:2.4,0.4848);
\draw[draw=black,fill=color1, postaction={pattern=north east lines},draw opacity=1] (axis cs:3,0) rectangle (axis cs:3.4,0.4438);
\draw[draw=black,fill=color1, postaction={pattern=north east lines},draw opacity=1] (axis cs:4,0) rectangle (axis cs:4.4,0.4418);
\draw[draw=black,fill=color1, postaction={pattern=north east lines},draw opacity=1] (axis cs:5,0) rectangle (axis cs:5.4,0.5179);
\draw[draw=black,fill=color1, postaction={pattern=north east lines},draw opacity=1] (axis cs:6,0) rectangle (axis cs:6.4,0.4669);
\draw[draw=black,fill=color1, postaction={pattern=north east lines},draw opacity=1] (axis cs:7,0) rectangle (axis cs:7.4,0.4493);
\draw[draw=black,fill=color1, postaction={pattern=north east lines},draw opacity=1] (axis cs:8,0) rectangle (axis cs:8.4,0.4685);
\draw[draw=black,fill=color1, postaction={pattern=north east lines},draw opacity=1] (axis cs:9,0) rectangle (axis cs:9.4,0.4504);
\draw[draw=black,fill=color1, postaction={pattern=north east lines},draw opacity=1] (axis cs:10,0) rectangle (axis cs:10.4,0.5359);
\draw[draw=black,fill=color1, postaction={pattern=north east lines},draw opacity=1] (axis cs:11,0) rectangle (axis cs:11.4,0.5289);
\draw[draw=black,fill=color1, postaction={pattern=north east lines},draw opacity=1] (axis cs:12,0) rectangle (axis cs:12.4,0.4377);
\draw[draw=black,fill=color1, postaction={pattern=north east lines},draw opacity=1] (axis cs:13,0) rectangle (axis cs:13.4,0.4878);
\draw[draw=black,fill=color1, postaction={pattern=north east lines},draw opacity=1] (axis cs:14,0) rectangle (axis cs:14.4,0.4479);
\draw[draw=black,fill=color1, postaction={pattern=north east lines},draw opacity=1] (axis cs:15,0) rectangle (axis cs:15.4,0.489);
\addplot [line width=1.08pt, white!26.0!black, forget plot]
table {%
-0.2 0.279085
-0.2 0.682835
};
\addplot [line width=1.08pt, white!26.0!black, forget plot]
table {%
0.8 0.5272875
0.8 0.7741
};
\addplot [line width=1.08pt, white!26.0!black, forget plot]
table {%
1.8 0.276888888888889
1.8 0.702447222222222
};
\addplot [line width=1.08pt, white!26.0!black, forget plot]
table {%
2.8 0.383844444444444
2.8 0.759705555555555
};
\addplot [line width=1.08pt, white!26.0!black, forget plot]
table {%
3.8 0.40109
3.8 0.7435125
};
\addplot [line width=1.08pt, white!26.0!black, forget plot]
table {%
4.8 0.442865
4.8 0.763335
};
\addplot [line width=1.08pt, white!26.0!black, forget plot]
table {%
5.8 0.36868
5.8 0.6457375
};
\addplot [line width=1.08pt, white!26.0!black, forget plot]
table {%
6.8 0.288340625
6.8 0.773875
};
\addplot [line width=1.08pt, white!26.0!black, forget plot]
table {%
7.8 0.4736875
7.8 0.7278
};
\addplot [line width=1.08pt, white!26.0!black, forget plot]
table {%
8.8 0.35023125
8.8 0.6555125
};
\addplot [line width=1.08pt, white!26.0!black, forget plot]
table {%
9.8 0.44836
9.8 0.73351
};
\addplot [line width=1.08pt, white!26.0!black, forget plot]
table {%
10.8 0.6067875
10.8 0.7931525
};
\addplot [line width=1.08pt, white!26.0!black, forget plot]
table {%
11.8 0.3317675
11.8 0.6814275
};
\addplot [line width=1.08pt, white!26.0!black, forget plot]
table {%
12.8 0.594195
12.8 0.7780025
};
\addplot [line width=1.08pt, white!26.0!black, forget plot]
table {%
13.8 0.412635
13.8 0.6770075
};
\addplot [line width=1.08pt, white!26.0!black, forget plot]
table {%
14.8 0.557655555555556
14.8 0.739561111111111
};
\addplot [line width=1.08pt, white!26.0!black, forget plot]
table {%
0.2 0.415895
0.2 0.4551025
};
\addplot [line width=1.08pt, white!26.0!black, forget plot]
table {%
1.2 0.4415925
1.2 0.4791
};
\addplot [line width=1.08pt, white!26.0!black, forget plot]
table {%
2.2 0.4725
2.2 0.496
};
\addplot [line width=1.08pt, white!26.0!black, forget plot]
table {%
3.2 0.4325975
3.2 0.4544025
};
\addplot [line width=1.08pt, white!26.0!black, forget plot]
table {%
4.2 0.4297975
4.2 0.455
};
\addplot [line width=1.08pt, white!26.0!black, forget plot]
table {%
5.2 0.4998
5.2 0.5397025
};
\addplot [line width=1.08pt, white!26.0!black, forget plot]
table {%
6.2 0.4549
6.2 0.4808025
};
\addplot [line width=1.08pt, white!26.0!black, forget plot]
table {%
7.2 0.44
7.2 0.459305
};
\addplot [line width=1.08pt, white!26.0!black, forget plot]
table {%
8.2 0.4531975
8.2 0.4857025
};
\addplot [line width=1.08pt, white!26.0!black, forget plot]
table {%
9.2 0.4386
9.2 0.462605
};
\addplot [line width=1.08pt, white!26.0!black, forget plot]
table {%
10.2 0.5111
10.2 0.5677125
};
\addplot [line width=1.08pt, white!26.0!black, forget plot]
table {%
11.2 0.5032925
11.2 0.562125
};
\addplot [line width=1.08pt, white!26.0!black, forget plot]
table {%
12.2 0.4194
12.2 0.451905
};
\addplot [line width=1.08pt, white!26.0!black, forget plot]
table {%
13.2 0.4753975
13.2 0.4999075
};
\addplot [line width=1.08pt, white!26.0!black, forget plot]
table {%
14.2 0.4362
14.2 0.4600075
};
\addplot [line width=1.08pt, white!26.0!black, forget plot]
table {%
15.2 0.472666666666667
15.2 0.506122222222222
};
\end{axis}

\end{tikzpicture}

%% file: plots/cap_prr_scalability.tex
\begin{tikzpicture}

\begin{axis}[
	legend columns=3,
width=.6*\FIGWIDTH,
height=5cm,
legend cell align={left},
legend style={fill opacity=1.8, draw opacity=1, text opacity=1, draw=white!80.0!black, anchor=south, at={(0.5,1.1)}},
tick align=outside,
tick pos=both,
x grid style={white!69.01960784313725!black},
xlabel={number of nodes},
xmin=-0.5, xmax=3.5,
xtick style={draw=black,color=black},
y grid style={white!69.01960784313725!black},
ylabel={PDR of secondary traffic},
ymin=0, ymax=0.923584719015893,
ytick style={draw=black,color=black},
xtick={0,1,2,3},
xticklabels={7,19,43,91}
]
\draw[draw=black,fill=color0,draw opacity=1] (axis cs:-0.4,0) rectangle (axis cs:-0.133333333333333,0.860571472138039);
\addlegendimage{ybar,ybar legend,draw=black,fill=color0,draw opacity=1};
\addlegendentry{QMA}

\draw[draw=black,fill=color0,draw opacity=1] (axis cs:0.6,0) rectangle (axis cs:0.866666666666667,0.779924013722482);
\draw[draw=black,fill=color0,draw opacity=1] (axis cs:1.6,0) rectangle (axis cs:1.86666666666667,0.699216152958521);
\draw[draw=black,fill=color0,draw opacity=1] (axis cs:2.6,0) rectangle (axis cs:2.86666666666667,0.680713367405209);
\draw[draw=black,fill=color1, postaction={pattern=north east lines},draw opacity=1] (axis cs:-0.133333333333333,0) rectangle (axis cs:0.133333333333333,0.400149031500838);
\addlegendimage{ybar,ybar legend,draw=black,fill=color1, postaction={pattern=north east lines},draw opacity=1};
\addlegendentry{slotted CSMA/CA}

\draw[draw=black,fill=color1, postaction={pattern=north east lines},draw opacity=1] (axis cs:0.866666666666667,0) rectangle (axis cs:1.13333333333333,0.487959507351658);
\draw[draw=black,fill=color1, postaction={pattern=north east lines},draw opacity=1] (axis cs:1.86666666666667,0) rectangle (axis cs:2.13333333333333,0.393504608401898);
\draw[draw=black,fill=color1, postaction={pattern=north east lines},draw opacity=1] (axis cs:2.86666666666667,0) rectangle (axis cs:3.13333333333333,0.378879922987698);
\draw[draw=black,fill=color2, postaction={pattern=dots},draw opacity=1] (axis cs:0.133333333333333,0) rectangle (axis cs:0.4,0.387168211041491);
\addlegendimage{ybar,ybar legend,draw=black,fill=color2, postaction={pattern=dots},draw opacity=1};
\addlegendentry{unslotted CSMA/CA}

\draw[draw=black,fill=color2, postaction={pattern=dots},draw opacity=1] (axis cs:1.13333333333333,0) rectangle (axis cs:1.4,0.447786702660384);
\draw[draw=black,fill=color2, postaction={pattern=dots},draw opacity=1] (axis cs:2.13333333333333,0) rectangle (axis cs:2.4,0.337583165542647);
\draw[draw=black,fill=color2, postaction={pattern=dots},draw opacity=1] (axis cs:3.13333333333333,0) rectangle (axis cs:3.4,0.317999261124528);
\addplot [line width=1.08pt, white!26.0!black, forget plot]
table {%
-0.266666666666667 0.844369906353559
-0.266666666666667 0.87960449430085
};
\addplot [line width=1.08pt, white!26.0!black, forget plot]
table {%
0.733333333333333 0.765552351389978
0.733333333333333 0.792441021800863
};
\addplot [line width=1.08pt, white!26.0!black, forget plot]
table {%
1.73333333333333 0.692824395986576
1.73333333333333 0.706255291749781
};
\addplot [line width=1.08pt, white!26.0!black, forget plot]
table {%
2.73333333333333 0.675490224997573
2.73333333333333 0.68546279594032
};
\addplot [line width=1.08pt, white!26.0!black, forget plot]
table {%
0 0.334060356193327
0 0.467574890695435
};
\addplot [line width=1.08pt, white!26.0!black, forget plot]
table {%
1 0.473640026424348
1 0.501314499522187
};
\addplot [line width=1.08pt, white!26.0!black, forget plot]
table {%
2 0.384842223868894
2 0.4026708122788
};
\addplot [line width=1.08pt, white!26.0!black, forget plot]
table {%
3 0.372839608846705
3 0.384834922823057
};
\addplot [line width=1.08pt, white!26.0!black, forget plot]
table {%
0.266666666666667 0.342752099327724
0.266666666666667 0.438737304200505
};
\addplot [line width=1.08pt, white!26.0!black, forget plot]
table {%
1.26666666666667 0.437413776288338
1.26666666666667 0.458352157859663
};
\addplot [line width=1.08pt, white!26.0!black, forget plot]
table {%
2.26666666666667 0.331755616218744
2.26666666666667 0.342670522124947
};
\addplot [line width=1.08pt, white!26.0!black, forget plot]
table {%
3.26666666666667 0.314157744776751
3.26666666666667 0.32226622774611
};
\end{axis}

\end{tikzpicture}

%% file: plots/cap_handshake_req.tex
\begin{tikzpicture}

\begin{axis}[
	width=.6*\FIGWIDTH,
	height=5cm,
	legend cell align={left},
	legend columns=3,
	legend style={fill opacity=1.8, draw opacity=1, text opacity=1, draw=white!80.0!black, anchor=south, at={(0.5,1.1)}},
legend cell align={left},
legend style={fill opacity=1.8, draw opacity=1, text opacity=1, draw=white!80.0!black},
tick align=outside,
tick pos=both,
x grid style={white!69.01960784313725!black},
xlabel={number of nodes},
xmin=-0.5, xmax=3.5,
xtick style={draw=black,color=black},
y grid style={white!69.01960784313725!black},
ylabel={successfull GTS-requests [\%]},
ymin=0, ymax=1.04193356954835,
ytick style={draw=black,color=black},
xtick={0,1,2,3},
xticklabels={7,19,43,91},
yticklabels={0,0,20,40,60,80,100},
]
\draw[draw=black,fill=color0,draw opacity=1] (axis cs:-0.4,0) rectangle (axis cs:-0.133333333333333,0.988902062547904);
\addlegendimage{ybar,ybar legend,draw=black,fill=color0,draw opacity=1};
\addlegendentry{QMA}

\draw[draw=black,fill=color0,draw opacity=1] (axis cs:0.6,0) rectangle (axis cs:0.866666666666667,0.955870538250101);
\draw[draw=black,fill=color0,draw opacity=1] (axis cs:1.6,0) rectangle (axis cs:1.86666666666667,0.934200622331084);
\draw[draw=black,fill=color0,draw opacity=1] (axis cs:2.6,0) rectangle (axis cs:2.86666666666667,0.925922179524328);
\draw[draw=black,fill=color1, postaction={pattern=north east lines},draw opacity=1] (axis cs:-0.133333333333333,0) rectangle (axis cs:0.133333333333333,0.73318904146271);
\addlegendimage{ybar,ybar legend,draw=black,fill=color1, postaction={pattern=north east lines},draw opacity=1};
\addlegendentry{slotted CSMA/CA}

\draw[draw=black,fill=color1, postaction={pattern=north east lines},draw opacity=1] (axis cs:0.866666666666667,0) rectangle (axis cs:1.13333333333333,0.810002116126518);
\draw[draw=black,fill=color1, postaction={pattern=north east lines},draw opacity=1] (axis cs:1.86666666666667,0) rectangle (axis cs:2.13333333333333,0.728966359496835);
\draw[draw=black,fill=color1, postaction={pattern=north east lines},draw opacity=1] (axis cs:2.86666666666667,0) rectangle (axis cs:3.13333333333333,0.708695308952764);
\draw[draw=black,fill=color2, postaction={pattern=dots},draw opacity=1] (axis cs:0.133333333333333,0) rectangle (axis cs:0.4,0.709111868635667);
\addlegendimage{ybar,ybar legend,draw=black,fill=color2, postaction={pattern=dots},draw opacity=1};
\addlegendentry{unslotted CSMA/CA}

\draw[draw=black,fill=color2, postaction={pattern=dots},draw opacity=1] (axis cs:1.13333333333333,0) rectangle (axis cs:1.4,0.772050970370484);
\draw[draw=black,fill=color2, postaction={pattern=dots},draw opacity=1] (axis cs:2.13333333333333,0) rectangle (axis cs:2.4,0.656217754574538);
\draw[draw=black,fill=color2, postaction={pattern=dots},draw opacity=1] (axis cs:3.13333333333333,0) rectangle (axis cs:3.4,0.625487266250686);
\addplot [line width=1.08pt, white!26.0!black, forget plot]
table {%
-0.266666666666667 0.98477368538475
-0.266666666666667 0.992317685284145
};
\addplot [line width=1.08pt, white!26.0!black, forget plot]
table {%
0.733333333333333 0.949428285349251
0.733333333333333 0.961684662539821
};
\addplot [line width=1.08pt, white!26.0!black, forget plot]
table {%
1.73333333333333 0.930789040891304
1.73333333333333 0.937682801887812
};
\addplot [line width=1.08pt, white!26.0!black, forget plot]
table {%
2.73333333333333 0.923074177873933
2.73333333333333 0.928584481966939
};
\addplot [line width=1.08pt, white!26.0!black, forget plot]
table {%
0 0.699850093736283
0 0.774794842506792
};
\addplot [line width=1.08pt, white!26.0!black, forget plot]
table {%
1 0.796123156363385
1 0.824854844980239
};
\addplot [line width=1.08pt, white!26.0!black, forget plot]
table {%
2 0.722591618626057
2 0.735393194964216
};
\addplot [line width=1.08pt, white!26.0!black, forget plot]
table {%
3 0.702199205030623
3 0.715591954312675
};
\addplot [line width=1.08pt, white!26.0!black, forget plot]
table {%
0.266666666666667 0.681686628792079
0.266666666666667 0.738927413269704
};
\addplot [line width=1.08pt, white!26.0!black, forget plot]
table {%
1.26666666666667 0.757837976515599
1.26666666666667 0.786096394885711
};
\addplot [line width=1.08pt, white!26.0!black, forget plot]
table {%
2.26666666666667 0.650797108238804
2.26666666666667 0.661854676244788
};
\addplot [line width=1.08pt, white!26.0!black, forget plot]
table {%
3.26666666666667 0.620340924745823
3.26666666666667 0.630569252585582
};
\end{axis}

\end{tikzpicture}